\documentclass[%
 reprint,
superscriptaddress,
citeautoscript
bibnotes,
amsmath,amssymb,
aps,
prb,
]{revtex4-2}
\usepackage{CJK}
\usepackage{color}
\usepackage{float}
\usepackage{graphicx}
\usepackage{dcolumn}
\usepackage{bm}
\usepackage{hyperref}

\begin{document}

\begin{CJK*}{GB}{}
\title{Effect of uncorrelated on-site Scalar Potential and Mass Disorder on Transport of Two-Dimensional Dirac Fermions}

\author{Arman Duha}
\affiliation{%
 Department of Physics, Oklahoma State University, Stillwater, Oklahoma 74078, USA
}%
\author{Mario F. Borunda}%
 \email{mario.borunda@okstate.edu}
\affiliation{%
 Department of Physics, Oklahoma State University, Stillwater, Oklahoma 74078, USA
}%

\maketitle
\end{CJK*}

\date{\today}

\begin{abstract}
We investigate the transport properties of massive Dirac fermions subjected to uncorrelated scalar potential disorder, and mass disorder. Using a finite difference method, the conductance is calculated for a wide variety of combinations of these two disorder strengths. By calculating the scaling of conductivity with system size we find that, depending on the combination, the system can have an insulating, scale invariant, and metallic behavior. We identify the critical values of these disorder strengths where the phase transitions occur. We study both the zero and nonzero average mass cases to examine the effect of scalar potential disorder on band gap. Our results suggest a suppression of the band gap by the scalar potential disorder.  
\end{abstract}

\maketitle


\section{Introduction}

Two-dimensional Dirac fermions realize a wide variety of symmetry classes when subjected to various types of disorder \citep{ludwig_integer_1994,bocquet2000disordered,bernard2002classification, evers_anderson_2008,ryu2010topological,PhysRevB.55.1142}.
Depending on the symmetry class, phase transitions due to Anderson localization \citep{anderson_absence_1958} vary considerably. These phase transitions include the metal-insulator transition (MIT) due to suppression of diffusion into localization and the insulator-insulator transition between separate localized phases \citep{PhysRevB.65.012506,PhysRevLett.101.127001,ponomarenko2011tunable,PhysRevB.85.195130,PhysRevB.106.104202}. 

Numerous studies \citep{nomura_topological_2007, bardarson_one-parameter_2007, tworzydlo_finite_2008, bardarson_absence_2010,  medvedyeva_effective_2010, medvedyeva_localization_2011, PhysRevB.79.075405,PhysRevB.97.125408,PhysRevResearch.4.L022035} have looked at the phase transitions for different symmetry classes of Dirac fermion systems. Graphene \citep{geim2007rise} and chiral p-wave superconductors \citep{kallin2009sr2ruo4,PhysRevLett.98.010506} are examples of some of the most studied such systems. In particular, disorder effects in transport properties are the subject of extensive investigation\citep{PhysRevB.108.064208,PhysRevB.108.064207,PhysRevB.101.115424,PhysRevX.4.041019,PhysRevB.104.045138,PhysRevB.101.121107}. If the disorder in graphene is random scalar potential, $V(x,y)$, which breaks the chiral and particle-hole symmetry, it falls in the symmetry class $AII$ \citep{ostrovsky_quantum_2007}. 
For the chiral $p$-wave superconductor system with vortex disorder, chiral and time-reversal symmetry are broken, but particle-hole symmetry remains invariant, which leads the system to symmetry class $BD$ \citep{bocquet2000disordered}.
Consequently, these two systems exhibit different types of phase transition with the variation of disorder strength.

\begin{figure}    
\includegraphics{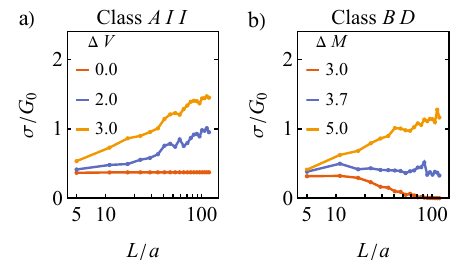}
\caption{Scaling of conductivity for different combinations of disorder. $\Delta V$ and $\Delta M$ are in the units of $\hbar v/a$ and $\hbar /va$, respectively.}
\label{two_classes}
\end{figure}

In clean graphene, the conductivity takes a scale-invariant critical value \citep{tworzydlo_sub-poissonian_2006, katsnelson2006zitterbewegung} of $\sigma_{c}=G_{0} / \pi$, where $G_{0}$ is the conductance quantum. 
When disorder $V$ is introduced (Class $AII$), this scale invariance is known to change into logarithmic scaling of conductivity \citep{nomura_topological_2007, bardarson_one-parameter_2007, tworzydlo_finite_2008} for sufficiently strong disorder (see Fig. \ref{two_classes}(a)). 
A random mass disorder, $M(x,y) = \bar M(x,y) + \delta M(x,y)$, can approximate a disordered chiral $p$-wave superconductor\citep{medvedyeva_effective_2010,PhysRevB.104.184201,PhysRevB.104.174205}, where $\bar M$ is the average mass and $\delta M$ is the random mass disorder. Consequently, instead of $V$, if the disorder is $M$, the system belongs to class $BD$. A very different phase transition is observed for this system \citep{medvedyeva_effective_2010}. 
Depending on the disorder's strength, the system can now be at an insulating or metallic phase. 
This insulator-to-metal transition is marked by a critical value of disorder strength for which the conductivity is scale-invariant, as indicated in Fig.~\ref{two_classes}(b). 

A more general case is when both scalar potential disorder $V$ and mass disorder $M$ are present. 
In a Dirac fermion system with both $V$ and $M$, all three symmetries are broken \citep{ludwig_integer_1994}, which puts the system into symmetry class $A$ \citep{bernard2002classification}. 
Such systems have been studied extensively \citep{ostrovsky_quantum_2007,titov_metal-insulator_2014,altland2002theories,nomura_quantum_2008,ludwig_integer_1994,hill2014scaling,PhysRevLett.126.076801} and includes examples like disordered $d$-wave superconductor \citep{altland2002theories}, and graphene on hexagonal boron nitride(h-BN) \citep{titov_metal-insulator_2014,decker2011local,dean2010boron}. 

\begin{figure*}
         \includegraphics{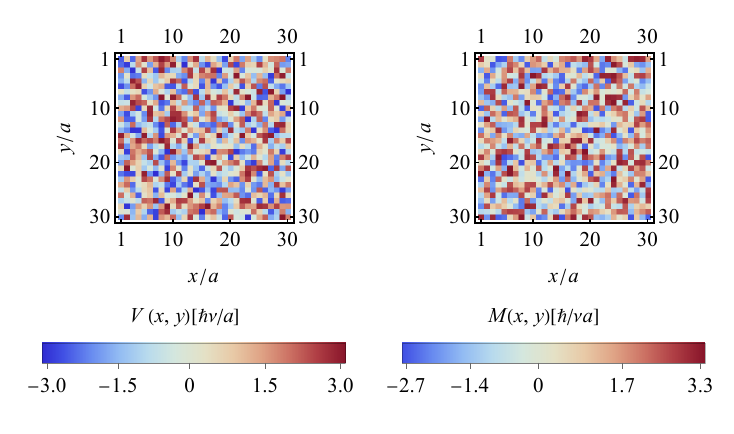}
         \caption{On-site random scalar potential $V(x,y) = \delta V$ (left) and random mass $M(x,y) = \bar{M} + \delta M$ (right). $\bar{M} = 0.3 \hbar/va$ and disorder strengths are $\Delta V = 3\hbar v/a$, $\Delta M = 3\hbar/va$.  }
         \label{contour}
\end{figure*}

In this manuscript, we numerically investigate the effect of disorder on conductivity when $\delta V$ and $\delta M$ are simultaneously present. 
The physics of localization for such a system depends sensitively on the range of the disorders. 
Intervalley scattering is significant when short-range correlated random disorder is present, and localization sets in \citep{nomura_quantum_2008, nomura_topological_2007}. 
For long-range correlated disorder, the conductivity flows to the quantum Hall critical point value, $\sigma_{xx}^{QHE}\approx 0.57\times 4e^2 $, for $\bar M = 0$ and to zero (insulating) for $\bar M\neq 0$. 
This insulating behavior is interpreted as a band gap opening induced by, for example, the proximity of graphene to h-BN \citep{titov_metal-insulator_2014}. On the other hand, it has been reported that uncorrelated onsite disorder can suppress the band gap and thereby induce a metallic state \citep{PhysRevB.94.195103,li2020band}. For graphene/h-BN systems, this occurs through the onsite Coulomb impurities restoring the inversion symmetry between neighboring carbons. 
It has been argued that this phenomenon of band gap suppression should be present in other van der Walls heterostructures as well \citep{PhysRevB.94.195103}.

This manuscript is organized as follows. 
In Sec. \ref{model}, we introduce the model and describe our system structure and method for calculating conductivity.
We choose both types of disorders, $\delta V(x,y)$ and $\delta M(x,y)$, to be on-site and spatially uncorrelated. 
$\delta V$ and $\delta M$ are distributed uniformly in the disorder strength interval, $(-\Delta V, \Delta V)$, and $(-\Delta M, \Delta M)$, respectively. 
In Sec. \ref{zero mbar}, we study the zero average mass ($\bar M=0$) case before dealing with the nonzero average mass ($\bar M \neq 0$) in Sec. \ref{nonzero mbar}. 
Under a wide variety of disorder strengths, we calculate the change in conductivity with system size using a finite difference approximation of the transfer matrix method \citep{tworzydlo_finite_2008}.
Our numerical results indicate that the system can behave as an insulator, scale-invariant with a critical conductivity value, and a diffusive metal depending on the average mass $\bar{M}$ and the disorder strengths. 
We look closer at the critical points, which separate two different phases, in Sec. \ref{critical}.
We conclude with the summary of our results in Sec. \ref{conclusion}.

\section{Model}\label{model}
We consider here a two-dimensional massive Dirac fermion Hamiltonian that satisfies the Dirac equation
\begin{equation}\label{eqn1}
    H\boldsymbol{\Psi}=E\boldsymbol{\Psi}, \hspace{0.5cm} H=v(\boldsymbol{\sigma} \cdot \boldsymbol{p})+V(x,y)+v^{2} M(x,y) \sigma_{z},
\end{equation}
where $\boldsymbol{\Psi}$ is the two-component spinor eigenstate, $E$ is the energy, and $v$ is the velocity of the Dirac fermion. 
The Pauli matrices ($\boldsymbol{\sigma}$) are given by the components 
\begin{align*}
\sigma_x =
\begin{pmatrix}0&1\\1&0\end{pmatrix},
\sigma_y =
\begin{pmatrix}0&-i\\i&0\end{pmatrix},
\sigma_z =
\begin{pmatrix}1&0\\0&-1\end{pmatrix}. 
\end{align*}
The scalar potential landscape is $V(x,y)$, and $M(x,y)$ is the local mass of the Dirac fermion. 
While the scalar potential $V$ breaks the chiral and particle-hole symmetries, the Dirac mass term $M$ breaks time-reversal symmetry, putting the system in class $A$. 

We model disorder by having at each lattice point a random value around the Fermi energy, i.e., $V(x,y)=E + \delta V(x,y)$, where $\delta V$ is the random scalar potential distributed uniformly in the interval $(-\Delta V, \Delta V)$. 
Similarly, a random mass fluctuation, $\delta M$, is introduced as $M(x,y)=\bar{M} + \delta M(x,y)$. 
Disorders are assigned randomly with a correlation length equal to the lattice constant $a$, illustrated in Fig.~\ref{contour}. 
These defects are realized through various physical and chemical processes. 
Any strain on the system and Coulomb impurities contribute to the scalar potential \citep{titov_metal-insulator_2014}. 
For multilayer systems, unprecedented precision of twist angle and interlayer distance \citep{liao2020precise, solis2020isothermal, brzhezinskaya2021engineering,uri2023superconductivity} has made it possible to control the interlayer coupling, which strongly affects the electronic structure and can result in a gap opening\citep{wang2016gaps,kerelsky2019maximized,PhysRevLett.129.116802,PhysRevB.103.075122} which corresponds to the mass term.  

In our calculations, we model the system around the Dirac point, $V= \delta V $. 
We consider four different values of average mass, $\bar{M}$, including the zero average mass case for bench-marking our results. 
The disorders, $\delta V$ and $\delta M$ are assigned at each lattice point independently from its neighboring lattice points, i.e., correlation length is equal to the lattice constant. 

We find the conductance of the system by mapping the Hamiltonian to a finite difference method (see Ref. \onlinecite{tworzydlo_finite_2008} for details) for solving the scattering by disorder problem of Dirac fermions. 
Our approach is the so-called staggered-fermion model, initially developed in lattice gauge theory \citep{PhysRevD.26.468, PhysRevLett.51.1815}, extensively used for calculating conductivity of different Dirac fermion systems \citep{tworzydlo_finite_2008,medvedyeva_effective_2010,bardarson_absence_2010,medvedyeva_localization_2011,borunda2011imaging,PhysRevB.88.125415,beenakker2023tangent,pacholski2021generalized}.

The geometry of the two-dimensional system we consider is of length $L=N_{x} a$ and width $W=N_{y} a$. 
All calculations assume periodic boundary conditions along the transverse direction with the system having an aspect ratio of $W/L=3$.  
The Hamiltonian and the boundary conditions are used to write the transfer matrix $\mathcal{T}$. 
The derivation of the transfer matrix $\mathcal{T}$ is presented in the Appendix for completeness. 
The transmission matrix $T$ was calculated from the evolution of eigenstates from one end of the system to the other via the transfer matrix. 
The conductance is calculated using the Landauer formula
\begin{align}
    G=G_0\sum_n T_n,
\end{align}
where $G_0$ is the conductance quantum and $T_n$ are the eigenvalues of the transmission matrix $T$. 
$ \langle G \rangle$ is the average of the conductance taken over many disorder realizations and is used to calculate the average conductivity, $\sigma = (L/W)\langle G\rangle$.

\section{Zero average mass $(\bar{M}=0)$}\label{zero mbar}
\begin{figure}[t]
      \includegraphics{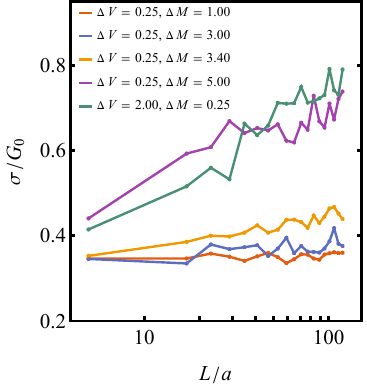}
      \caption{Scaling of conductivity behavior for solo disorder (either $\Delta V$ or $\Delta M$) is recovered in the presence of dual disorder ($\Delta V$ and $\Delta M$) for vanishing value of one.}
      \label{M0_scaling}  
\end{figure}

The scaling behavior of conductivity in the presence of one type of disorder (either scalar potential or mass) while the other type is absent is well known \citep{tworzydlo_finite_2008,medvedyeva_effective_2010,PhysRevB.104.184201,PhysRevB.104.174205,abergel2010properties,PhysRevLett.102.126802,PhysRevB.81.241404}. 
To investigate the combined effect of the two disorders, we start with two limiting cases, i.e., small mass fluctuations ($\Delta M \approx 0$) with $\Delta V>0$, and small potential fluctuations ($\Delta V \approx 0$) with $\Delta M>0$. 
For small $\Delta M$ and nonzero $\Delta V$, the conductivity scales towards that of a diffusive metal, as shown by the green line in Fig.~\ref{M0_scaling}. 
These results agree with Ref.~\onlinecite{tworzydlo_finite_2008}. 

On the other hand, for small $\Delta V$ and nonzero $\Delta M$, two phases are observed at $\bar M=0$ \citep{PhysRevB.65.012506, PhysRevLett.101.127001}, separated by a tricritical point, $\Delta M^{*}$ \citep{medvedyeva_effective_2010}. 
The tricritical point is defined as the value of mass disorder such that
for $\Delta M<\Delta M^{*}$ the system approaches the scale-invariant critical conductivity and $\Delta M>\Delta M^{*}$ drives the system to a diffusive metal phase. 
We identify this behavior in Fig. \ref{M0_scaling} for a small $\Delta V$ ($= 0.25 \hbar v/a$). The system is driven to a diffusive metal phase around $\Delta M^{*} \approx 3.2\hbar/va$ and remains metallic for still higher values of $\Delta M$. 
This result agrees with previous numerical results \citep{medvedyeva_effective_2010,PhysRevLett.102.126802,PhysRevB.81.241404} found for the $\Delta V=0$ case.

\begin{figure}[b]
      \includegraphics{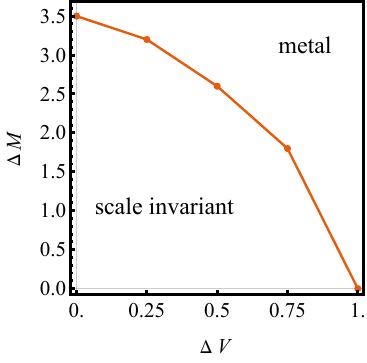}
      \caption{Phase diagram for the $\bar{M}=0$ case. Data points are the critical mass disorder values, $\Delta M^*$, that separate the scale invariant phase from the metallic phase at the corresponding $\Delta V$ }
      \label{M0_phase}  
\end{figure}

In Fig. \ref{M0_phase}, we present the phase diagram by numerically calculating $\Delta M^*$ for different $\Delta V$ values. 
$\Delta M^*$ decreases with increasing $\Delta V$ and eventually becomes zero. 
We define this value of $\Delta V$, for which $\Delta M^*$ reaches zero, as the critical scalar potential disorder, $\Delta V^*$. for $\Delta V>\Delta V^*$, the system loses access to the scale invariant phase and must remain metallic regardless of the $\Delta M$ value. 
For the massless case here, $\Delta V^{*}=1 \hbar v/a$, and as we will see in the next section, $\Delta V^*$ increases with increasing average mass $\bar{M}$. 
    
\begin{figure*}
     
         \includegraphics{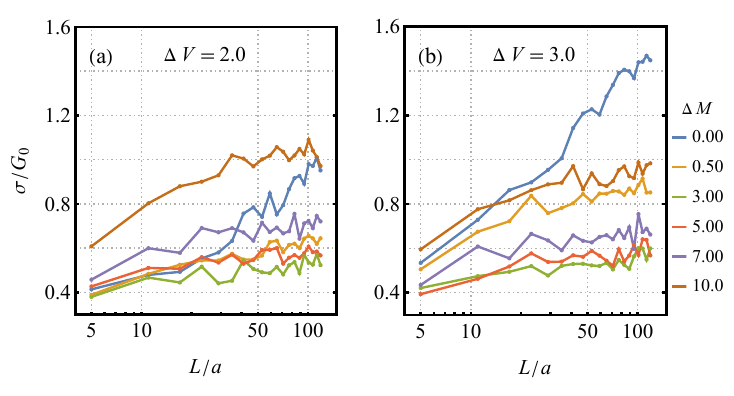}
         \caption{Scaling of conductivity $\sigma$ for different $\Delta M$ at $\bar M=0$. Two $\Delta V$ cases are demonstrated for both of which $\sigma$ starts decreasing at first with increasing $\Delta M$ and reaches a minimum value before increasing with $\Delta M$. }
         \label{two_dV}
     
\end{figure*}

We now consider the case when both $\Delta V$ and $\Delta M$ are comparable in strength. 
We consider a wide range of $\Delta M$ values for two fixed values of $\Delta V$ to examine the conductivity under their combined effect.
Like the lone $\Delta V$ disorder case, the conductivity always increases logarithmically with system size $L$. 
This increase occurs regardless of the disorder strength $\Delta M$.
However, compared to the lone $\Delta V$ case, the conductivity curve is significantly flattened, i.e., conductivity increases much slower with system size when $\Delta V$ and $\Delta M$ are present simultaneously.

Figure \ref{two_dV} also indicates that increasing $\Delta M$ initially decreases the overall conductivity, presumably due to some competing effect between $\Delta V$ and $\Delta M$. However, after reaching a minimum value, the conductivity increases with $\Delta M$. This increase occurs only when $\Delta M$ is large enough to be the dominant disorder such that it increases the conductivity, much like when $\Delta M$ is the only disorder.

\begin{figure*}
     
         \includegraphics{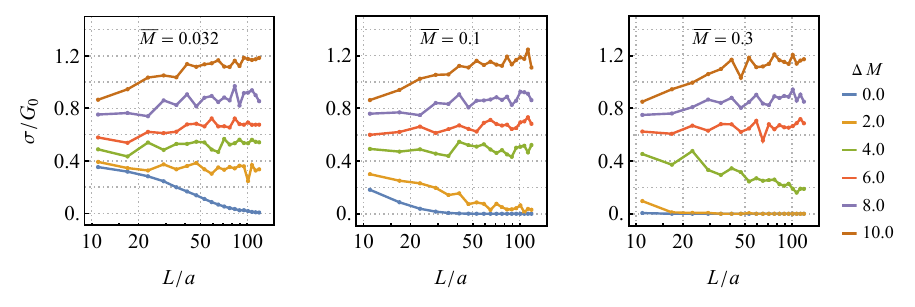}
         \caption{Scaling of conductivity $\sigma$ for different $\Delta M$ in the $\Delta V<\Delta V^*$ case. The system exhibits insulating, scale invariant, and metallic behavior. $\Delta M^*$ increases with increasing $\bar M$. In all three cases $\Delta V = 1\hbar v/a$ }
         \label{below_dV}
     
\end{figure*}

\section{Nonzero average mass $(\bar{M} \neq 0)$}\label{nonzero mbar}
So far, we have considered the zero average Dirac mass case $(\bar{M}=0)$. 
We now examine the nonzero $\bar{M}$ case in the presence of $\Delta V$ and $\Delta M$. 
For correlated disorder, A nonzero $\bar M$ is known to drive the system to an insulator. 
We also identify such an insulating phase for uncorrelated $\Delta M$, but this phase is now controlled by both $\Delta V$ and $\Delta M$. 
Only for $\Delta V$ and $\Delta M$ values smaller than particular critical values, $\Delta V^{*}$ and $\Delta M^{*}$, does the insulating phase exist. 
We discuss these critical values in Sec. \ref{critical}.

\begin{figure*}
     
         \includegraphics{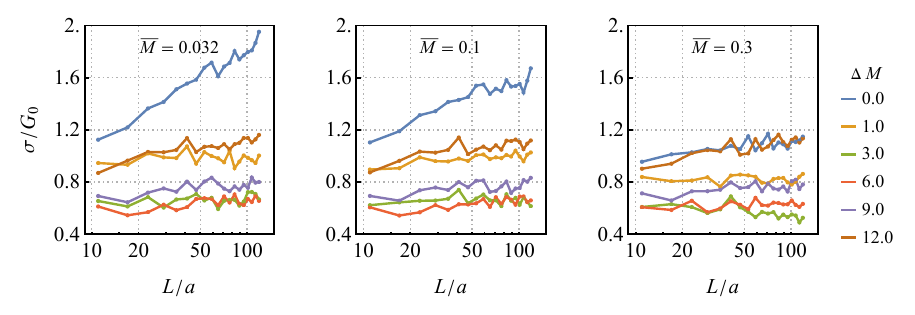}
         \caption{Scaling of conductivity $\sigma$ for different $\Delta M$ in the $\Delta V>\Delta V^*$ case. The system exhibits only metallic behavior. Similar to the $\bar M=0$ case, $\sigma$ starts decreasing at first with increasing $\Delta M$ and reaches a minimum value before increasing with increasing $\Delta M$. In all three cases $\Delta V = 5\hbar v/a$  }
         \label{above_dV}
     
\end{figure*}

When $\Delta V<\Delta V^{*}$, various conductivity scaling behaviors are observed. 
Figure \ref{below_dV} shows that as we increase the $\Delta M$ values, the system switches from insulating to a scale invariant conductivity and finally to a metallic phase. 
We identify the $\Delta M$ value at which the system transitions from insulating behavior to scale invariant conductivity as the critical mass disorder  $\equiv \Delta M^*$. 
This critical value depends on $\bar{M}$ and $\Delta V$. 
If the system's average mass $\bar M$ increases, so does the required mass disorder, $\Delta M^*$, to lift the system from the insulating phase. However, once the system reaches the metallic phase at $\Delta M$ values higher than its corresponding $\Delta M^*$, the conductivity scaling remains the same for different $\bar M$ values, as seen in Fig.~\ref{below_dV}.  

There is no insulating phase for $\Delta V>\Delta V^*$. 
The variation of conductivity scaling with $\Delta M$ is similar to that of the $\bar M=0$ case. 
In other words, the band gap $(\bar M)$ is suppressed by introducing $\Delta V$, and for a sufficiently strong value, the conductivity is much like the closed band ($\bar M=0$) case. 
This is indicated in Fig.~\ref{above_dV}, which shows that, like the $\bar M=0$ case, conductivity first decreases with $\Delta M$ and reaches a minimum value before increasing with larger values of $\Delta M$. 
This result agrees with the results found in Ref. \onlinecite{PhysRevB.94.195103} and \onlinecite{,li2020band} that the band gap is suppressed by uncorrelated onsite scalar potential.

\begin{figure*}
     
         \includegraphics{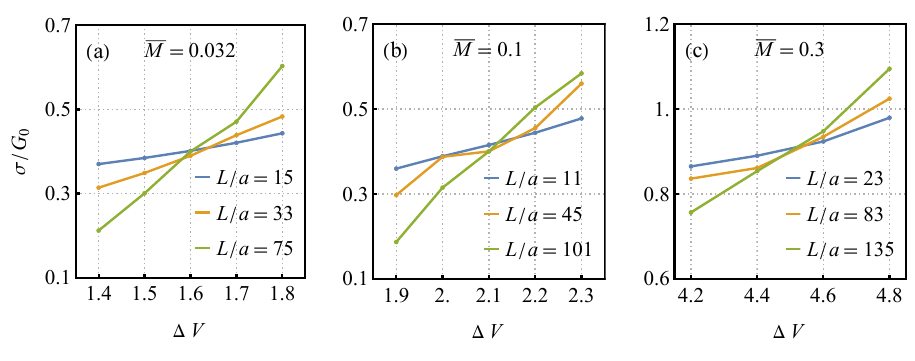}
         \caption{Change of conductivity $\sigma$ with $\Delta V$. $\Delta V^*$ is calculated by identifying the value of $\Delta V$ at which the conductivity becomes scale invariant. The required $\Delta V^*$ value increases with $\bar M$}
         \label{critical_dV}
     
\end{figure*}
\begin{figure*}
     
         \includegraphics{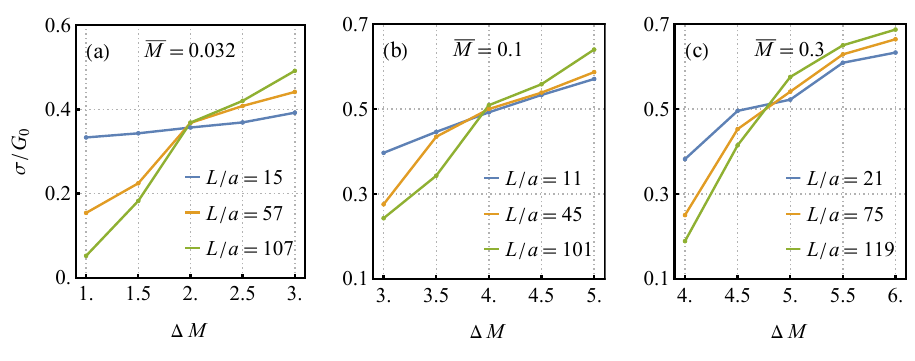}
         \caption{Change of conductivity $\sigma$ with $\Delta M$. $\Delta M^*$ is calculated by identifying the value of $\Delta M$ at which the conductivity becomes scale invariant. $\Delta V$ is fixed at $\Delta V=1 \hbar v/a$ }
         \label{critical_dM}
     
\end{figure*}

\section{Critical points}\label{critical}

In Sec. \ref{nonzero mbar}, the transitions of scaling behavior (from insulating to scale invariant and metallic) are demonstrated without specifying the transition points. 
We now focus on numerically evaluating the critical points, $\Delta V^{*}$ and $\Delta M^{*}$, at which these transitions occur.
As mentioned, the insulating phase can only exist for $\Delta V$ values below a critical value, $\Delta V^{*}$. 
We interpret $\Delta V^*$ as the onsite potential disorder width necessary for quelling the band gap to zero. 
$\Delta V^*$ thus depends on $\bar M$ (band gap) i.e., $\Delta V^*=\Delta V^*(\bar M)$. In Fig. \ref{critical_dV}, we identify the $\Delta V^{*}(\bar{M})$ values by calculating the scale dependence of conductivity for different $\Delta V$ at $\Delta M=0$.
As expected, a more significant value of $\bar{M}$ requires a larger $\Delta V^{*}$ to change the conductivity scaling from insulating to scale invariant.

The critical scalar potential disorder strength, $\Delta V^{*}$, acts as a boundary above which the insulating phase disappears, but for $\Delta V<\Delta V^*$, the system can have all three scaling behaviors. 
Precisely which scale dependence the system will exhibit for a particular $\Delta V$ depends on the value of the mass disorder strength, $\Delta M$. 
We illustrate this dependence in Fig. \ref{below_dV} where we notice that the system is insulating for $\Delta M=0$ and reaches a scale-invariant conductivity as we increase $\Delta M$. 
We identify the $\Delta M^*$ values by calculating the scale dependence of conductivity for three different $\bar M$ values with fixed $\Delta V=1 \hbar v/a$. 
The results are shown in Fig. \ref{critical_dM}. 

After reaching the scale invariant conductivity at $\Delta M = \Delta M^*$, the conductivity increase with system size for $\Delta M$ values larger than $\Delta M^*$ is relatively slow. 
This trend is shown in Fig.~\ref{below_dV}. 
It is part of the general observation that when $\Delta V$ and $\Delta M$ are of comparable strength, and the system is in a metallic phase, the system is a poor conductor because the conductivity increase rate with system size $L$ is sluggish. 
This is also seen in Fig.~\ref{above_dV} where we notice that the conductivity is lowest and has 
the flattest scaling when the disorder strengths $\Delta V$ and $\Delta M$ are similar. 

\section{Conclusion}\label{conclusion}

In this manuscript, we have studied the transport properties of a massive Dirac fermion in the simultaneous presence of scalar potential disorder $\Delta V$ and mass disorder $\Delta M$. 
Our numerical calculations use the real space tight-binding model on a lattice with on-site uncorrelated disorder developed by Tworzyd\l{}o {\em et at.} \citep{tworzydlo_finite_2008,medvedyeva_effective_2010}. 
We study three different average masses, $\bar M$, which is interpreted as the band gap. In all three cases, despite the band gap, we identify that a critical $\Delta V^*(\bar M)$ exists above which the system can no longer be an insulator for any $\Delta M$. 
The results support the idea of band gap suppression by on-site coulomb potential discussed in Ref. \onlinecite{PhysRevB.94.195103} and \onlinecite{li2020band}. 
For $\Delta V<\Delta V^*$, the system can be in an insulating or metallic phase, depending on the $\Delta M$ value. 
As $\Delta M$ increases, the system exhibits an insulator-to-metal transition at a critical value $\Delta M^*(\bar M, \Delta V)$. 
We have numerically estimated the critical values, $\Delta V^*$, and $\Delta M^*$, for different $\bar M$ values. 
Our work demonstrates the interdependent way different types of disorders can affect the phases accessible to a massive Dirac fermion system. 

\begin{acknowledgments}
The calculations were performed in the PETE system of the High Performance Computing Center at Oklahoma State University, NSF grant no. OAC-1531128.
\end{acknowledgments}
\appendix*
\section{Transfer Matrix}
The Dirac equation given in Eq. (\ref{eqn1}) can be rearranged as,
\[
\partial_x\boldsymbol{\Psi}=\left(-i\sigma_z\partial_y-i\sigma _x \frac{V}{\hbar v}-\frac{v M}{\hbar} \sigma_y\right)\boldsymbol{\Psi}
\]
Using the discretization method of Ref. \onlinecite{tworzydlo_finite_2008}, the discretized Dirac equation is expressed by

\[
\frac{1}{2a}\mathcal{J}(\boldsymbol{\Psi}_{m+1}-\boldsymbol{\Psi}_m)
\]
\[= \left(-\frac{i}{2a}\sigma_z\mathcal{K}-\frac{i}{4\hbar v}\sigma_x\mathcal{V}^{(m)}-\frac{v}{4\hbar} \sigma_y\mathcal{M}^{(m)}\right)(\boldsymbol{\Psi}_m+\boldsymbol{\Psi}_{m+1})
\]

where $\mathcal{J}$, $\mathcal{K}$, $\mathcal{V}^{(m)}$, and $\mathcal{M}^{(m)}$ are matrices with the following nonzero elements:
\[
\mathcal{J}_{n,n}= 1, \hspace{0.5 cm}\mathcal{J}_{n,n+1}=\mathcal{J}_{n,n-1}=\frac{1}{2},
\]

\[
\mathcal{K}_{n,n+1}=\frac{1}{2}, \hspace{0.5 cm}\mathcal{K}_{n,n-1}=-\frac{1}{2},
\]
\[
\mathcal{V}_{n,n}^{(m)}=\frac{1}{2}(V_{m,n}+V_{m,n-1}), \hspace{0.5 cm} \mathcal{V}_{n,n+1}^{(m)}=\frac{1}{2}V_{m,n},
\]
\[
\mathcal{V}_{n,n-1}^{(m)}=\frac{1}{2}V_{m,n-1},
\]
\[
\mathcal{M}_{n,n}^{(m)}=\frac{1}{2}(M_{m,n}+M_{m,n-1}), \hspace{0.5 cm} \mathcal{M}_{n,n+1}^{(m)}=\frac{1}{2}M_{m,n},
\]
\[
\mathcal{M}_{n,n-1}^{(m)}=\frac{1}{2}M_{m,n-1}.
\]

The transfer matrix, $\mathcal{T}_m$, which is defined as,
\[
\boldsymbol{\Psi}_{m+1}=\mathcal{T}_m\boldsymbol{\Psi}_m
\]
is then given by
\[
\mathcal{T}_m=\frac{\left(\mathcal{J}-i\sigma J\mathcal{K}-\frac{a}{2\hbar v}i\sigma_x \mathcal{V}^{(m)}-\frac{va}{2\hbar}\sigma_y\mathcal{M}^{(m)}\right)}{\left(\mathcal{J}+i\sigma J\mathcal{K}+\frac{a}{2\hbar v}i\sigma_x \mathcal{V}^{(m)}+\frac{va}{2\hbar}\sigma_y\mathcal{M}^{(m)}\right)}
\]

\bibliography{sfm}

\begin{thebibliography}{61}%
\makeatletter
\providecommand \@ifxundefined [1]{%
 \@ifx{#1\undefined}
}%
\providecommand \@ifnum [1]{%
 \ifnum #1\expandafter \@firstoftwo
 \else \expandafter \@secondoftwo
 \fi
}%
\providecommand \@ifx [1]{%
 \ifx #1\expandafter \@firstoftwo
 \else \expandafter \@secondoftwo
 \fi
}%
\providecommand \natexlab [1]{#1}%
\providecommand \enquote  [1]{``#1''}%
\providecommand \bibnamefont  [1]{#1}%
\providecommand \bibfnamefont [1]{#1}%
\providecommand \citenamefont [1]{#1}%
\providecommand \href@noop [0]{\@secondoftwo}%
\providecommand \href [0]{\begingroup \@sanitize@url \@href}%
\providecommand \@href[1]{\@@startlink{#1}\@@href}%
\providecommand \@@href[1]{\endgroup#1\@@endlink}%
\providecommand \@sanitize@url [0]{\catcode `\\12\catcode `\$12\catcode
  `\&12\catcode `\#12\catcode `\^12\catcode `\_12\catcode `\%12\relax}%
\providecommand \@@startlink[1]{}%
\providecommand \@@endlink[0]{}%
\providecommand \url  [0]{\begingroup\@sanitize@url \@url }%
\providecommand \@url [1]{\endgroup\@href {#1}{\urlprefix }}%
\providecommand \urlprefix  [0]{URL }%
\providecommand \Eprint [0]{\href }%
\providecommand \doibase [0]{https://doi.org/}%
\providecommand \selectlanguage [0]{\@gobble}%
\providecommand \bibinfo  [0]{\@secondoftwo}%
\providecommand \bibfield  [0]{\@secondoftwo}%
\providecommand \translation [1]{[#1]}%
\providecommand \BibitemOpen [0]{}%
\providecommand \bibitemStop [0]{}%
\providecommand \bibitemNoStop [0]{.\EOS\space}%
\providecommand \EOS [0]{\spacefactor3000\relax}%
\providecommand \BibitemShut  [1]{\csname bibitem#1\endcsname}%
\let\auto@bib@innerbib\@empty
\bibitem [{\citenamefont {Ludwig}\ \emph {et~al.}(1994)\citenamefont {Ludwig},
  \citenamefont {Fisher}, \citenamefont {Shankar},\ and\ \citenamefont
  {Grinstein}}]{ludwig_integer_1994}%
  \BibitemOpen
  \bibfield  {author} {\bibinfo {author} {\bibfnamefont {A.~W.~W.}\
  \bibnamefont {Ludwig}}, \bibinfo {author} {\bibfnamefont {M.~P.~A.}\
  \bibnamefont {Fisher}}, \bibinfo {author} {\bibfnamefont {R.}~\bibnamefont
  {Shankar}},\ and\ \bibinfo {author} {\bibfnamefont {G.}~\bibnamefont
  {Grinstein}},\ }\bibfield  {title} {\bibinfo {title} {Integer quantum {Hall}
  transition: {An} alternative approach and exact results},\ }\href
  {https://doi.org/10.1103/PhysRevB.50.7526} {\bibfield  {journal} {\bibinfo
  {journal} {Physical Review B}\ }\textbf {\bibinfo {volume} {50}},\ \bibinfo
  {pages} {7526} (\bibinfo {year} {1994})},\ \bibinfo {note} {publisher:
  American Physical Society}\BibitemShut {NoStop}%
\bibitem [{\citenamefont {Bocquet}\ \emph {et~al.}(2000)\citenamefont
  {Bocquet}, \citenamefont {Serban},\ and\ \citenamefont
  {Zirnbauer}}]{bocquet2000disordered}%
  \BibitemOpen
  \bibfield  {author} {\bibinfo {author} {\bibfnamefont {M.}~\bibnamefont
  {Bocquet}}, \bibinfo {author} {\bibfnamefont {D.}~\bibnamefont {Serban}},\
  and\ \bibinfo {author} {\bibfnamefont {M.}~\bibnamefont {Zirnbauer}},\
  }\bibfield  {title} {\bibinfo {title} {Disordered 2d quasiparticles in class
  d: Dirac fermions with random mass, and dirty superconductors},\ }\href
  {https://doi.org/https://doi.org/10.1016/S0550-3213(00)00208-X} {\bibfield
  {journal} {\bibinfo  {journal} {Nuclear Physics B}\ }\textbf {\bibinfo
  {volume} {578}},\ \bibinfo {pages} {628} (\bibinfo {year}
  {2000})}\BibitemShut {NoStop}%
\bibitem [{\citenamefont {Bernard}\ and\ \citenamefont
  {LeClair}(2002)}]{bernard2002classification}%
  \BibitemOpen
  \bibfield  {author} {\bibinfo {author} {\bibfnamefont {D.}~\bibnamefont
  {Bernard}}\ and\ \bibinfo {author} {\bibfnamefont {A.}~\bibnamefont
  {LeClair}},\ }\bibfield  {title} {\bibinfo {title} {A classification of 2d
  random dirac fermions},\ }\href
  {https://iopscience.iop.org/article/10.1088/0305-4470/35/11/303/meta}
  {\bibfield  {journal} {\bibinfo  {journal} {Journal of Physics A:
  Mathematical and General}\ }\textbf {\bibinfo {volume} {35}},\ \bibinfo
  {pages} {2555} (\bibinfo {year} {2002})}\BibitemShut {NoStop}%
\bibitem [{\citenamefont {Evers}\ and\ \citenamefont
  {Mirlin}(2008)}]{evers_anderson_2008}%
  \BibitemOpen
  \bibfield  {author} {\bibinfo {author} {\bibfnamefont {F.}~\bibnamefont
  {Evers}}\ and\ \bibinfo {author} {\bibfnamefont {A.~D.}\ \bibnamefont
  {Mirlin}},\ }\bibfield  {title} {\bibinfo {title} {Anderson transitions},\
  }\href {https://doi.org/10.1103/RevModPhys.80.1355} {\bibfield  {journal}
  {\bibinfo  {journal} {Reviews of Modern Physics}\ }\textbf {\bibinfo {volume}
  {80}},\ \bibinfo {pages} {1355} (\bibinfo {year} {2008})},\ \bibinfo {note}
  {publisher: American Physical Society}\BibitemShut {NoStop}%
\bibitem [{\citenamefont {Ryu}\ \emph {et~al.}(2010)\citenamefont {Ryu},
  \citenamefont {Schnyder}, \citenamefont {Furusaki},\ and\ \citenamefont
  {Ludwig}}]{ryu2010topological}%
  \BibitemOpen
  \bibfield  {author} {\bibinfo {author} {\bibfnamefont {S.}~\bibnamefont
  {Ryu}}, \bibinfo {author} {\bibfnamefont {A.~P.}\ \bibnamefont {Schnyder}},
  \bibinfo {author} {\bibfnamefont {A.}~\bibnamefont {Furusaki}},\ and\
  \bibinfo {author} {\bibfnamefont {A.~W.}\ \bibnamefont {Ludwig}},\ }\bibfield
   {title} {\bibinfo {title} {Topological insulators and superconductors:
  tenfold way and dimensional hierarchy},\ }\href
  {https://iopscience.iop.org/article/10.1088/1367-2630/12/6/065010/meta}
  {\bibfield  {journal} {\bibinfo  {journal} {New Journal of Physics}\ }\textbf
  {\bibinfo {volume} {12}},\ \bibinfo {pages} {065010} (\bibinfo {year}
  {2010})}\BibitemShut {NoStop}%
\bibitem [{\citenamefont {Altland}\ and\ \citenamefont
  {Zirnbauer}(1997)}]{PhysRevB.55.1142}%
  \BibitemOpen
  \bibfield  {author} {\bibinfo {author} {\bibfnamefont {A.}~\bibnamefont
  {Altland}}\ and\ \bibinfo {author} {\bibfnamefont {M.~R.}\ \bibnamefont
  {Zirnbauer}},\ }\bibfield  {title} {\bibinfo {title} {Nonstandard symmetry
  classes in mesoscopic normal-superconducting hybrid structures},\ }\href
  {https://doi.org/10.1103/PhysRevB.55.1142} {\bibfield  {journal} {\bibinfo
  {journal} {Phys. Rev. B}\ }\textbf {\bibinfo {volume} {55}},\ \bibinfo
  {pages} {1142} (\bibinfo {year} {1997})}\BibitemShut {NoStop}%
\bibitem [{\citenamefont {Anderson}(1958)}]{anderson_absence_1958}%
  \BibitemOpen
  \bibfield  {author} {\bibinfo {author} {\bibfnamefont {P.~W.}\ \bibnamefont
  {Anderson}},\ }\bibfield  {title} {\bibinfo {title} {Absence of {Diffusion}
  in {Certain} {Random} {Lattices}},\ }\href
  {https://doi.org/10.1103/PhysRev.109.1492} {\bibfield  {journal} {\bibinfo
  {journal} {Physical Review}\ }\textbf {\bibinfo {volume} {109}},\ \bibinfo
  {pages} {1492} (\bibinfo {year} {1958})},\ \bibinfo {note} {publisher:
  American Physical Society}\BibitemShut {NoStop}%
\bibitem [{\citenamefont {Chalker}\ \emph {et~al.}(2001)\citenamefont
  {Chalker}, \citenamefont {Read}, \citenamefont {Kagalovsky}, \citenamefont
  {Horovitz}, \citenamefont {Avishai},\ and\ \citenamefont
  {Ludwig}}]{PhysRevB.65.012506}%
  \BibitemOpen
  \bibfield  {author} {\bibinfo {author} {\bibfnamefont {J.~T.}\ \bibnamefont
  {Chalker}}, \bibinfo {author} {\bibfnamefont {N.}~\bibnamefont {Read}},
  \bibinfo {author} {\bibfnamefont {V.}~\bibnamefont {Kagalovsky}}, \bibinfo
  {author} {\bibfnamefont {B.}~\bibnamefont {Horovitz}}, \bibinfo {author}
  {\bibfnamefont {Y.}~\bibnamefont {Avishai}},\ and\ \bibinfo {author}
  {\bibfnamefont {A.~W.~W.}\ \bibnamefont {Ludwig}},\ }\bibfield  {title}
  {\bibinfo {title} {Thermal metal in network models of a disordered
  two-dimensional superconductor},\ }\href
  {https://doi.org/10.1103/PhysRevB.65.012506} {\bibfield  {journal} {\bibinfo
  {journal} {Phys. Rev. B}\ }\textbf {\bibinfo {volume} {65}},\ \bibinfo
  {pages} {012506} (\bibinfo {year} {2001})}\BibitemShut {NoStop}%
\bibitem [{\citenamefont {Kagalovsky}\ and\ \citenamefont
  {Nemirovsky}(2008)}]{PhysRevLett.101.127001}%
  \BibitemOpen
  \bibfield  {author} {\bibinfo {author} {\bibfnamefont {V.}~\bibnamefont
  {Kagalovsky}}\ and\ \bibinfo {author} {\bibfnamefont {D.}~\bibnamefont
  {Nemirovsky}},\ }\bibfield  {title} {\bibinfo {title} {Universal critical
  exponent in class d superconductors},\ }\href
  {https://doi.org/10.1103/PhysRevLett.101.127001} {\bibfield  {journal}
  {\bibinfo  {journal} {Phys. Rev. Lett.}\ }\textbf {\bibinfo {volume} {101}},\
  \bibinfo {pages} {127001} (\bibinfo {year} {2008})}\BibitemShut {NoStop}%
\bibitem [{\citenamefont {Ponomarenko}\ \emph {et~al.}(2011)\citenamefont
  {Ponomarenko}, \citenamefont {Geim}, \citenamefont {Zhukov}, \citenamefont
  {Jalil}, \citenamefont {Morozov}, \citenamefont {Novoselov}, \citenamefont
  {Grigorieva}, \citenamefont {Hill}, \citenamefont {Cheianov}, \citenamefont
  {Fal'ko} \emph {et~al.}}]{ponomarenko2011tunable}%
  \BibitemOpen
  \bibfield  {author} {\bibinfo {author} {\bibfnamefont {L.}~\bibnamefont
  {Ponomarenko}}, \bibinfo {author} {\bibfnamefont {A.}~\bibnamefont {Geim}},
  \bibinfo {author} {\bibfnamefont {A.}~\bibnamefont {Zhukov}}, \bibinfo
  {author} {\bibfnamefont {R.}~\bibnamefont {Jalil}}, \bibinfo {author}
  {\bibfnamefont {S.}~\bibnamefont {Morozov}}, \bibinfo {author} {\bibfnamefont
  {K.}~\bibnamefont {Novoselov}}, \bibinfo {author} {\bibfnamefont
  {I.}~\bibnamefont {Grigorieva}}, \bibinfo {author} {\bibfnamefont
  {E.}~\bibnamefont {Hill}}, \bibinfo {author} {\bibfnamefont {V.}~\bibnamefont
  {Cheianov}}, \bibinfo {author} {\bibfnamefont {V.}~\bibnamefont {Fal'ko}},
  \emph {et~al.},\ }\bibfield  {title} {\bibinfo {title} {Tunable
  metal--insulator transition in double-layer graphene heterostructures},\
  }\href {https://www.nature.com/articles/nphys2114} {\bibfield  {journal}
  {\bibinfo  {journal} {Nature Physics}\ }\textbf {\bibinfo {volume} {7}},\
  \bibinfo {pages} {958} (\bibinfo {year} {2011})}\BibitemShut {NoStop}%
\bibitem [{\citenamefont {K\"onig}\ \emph {et~al.}(2012)\citenamefont
  {K\"onig}, \citenamefont {Ostrovsky}, \citenamefont {Protopopov},\ and\
  \citenamefont {Mirlin}}]{PhysRevB.85.195130}%
  \BibitemOpen
  \bibfield  {author} {\bibinfo {author} {\bibfnamefont {E.~J.}\ \bibnamefont
  {K\"onig}}, \bibinfo {author} {\bibfnamefont {P.~M.}\ \bibnamefont
  {Ostrovsky}}, \bibinfo {author} {\bibfnamefont {I.~V.}\ \bibnamefont
  {Protopopov}},\ and\ \bibinfo {author} {\bibfnamefont {A.~D.}\ \bibnamefont
  {Mirlin}},\ }\bibfield  {title} {\bibinfo {title} {Metal-insulator transition
  in two-dimensional random fermion systems of chiral symmetry classes},\
  }\href {https://doi.org/10.1103/PhysRevB.85.195130} {\bibfield  {journal}
  {\bibinfo  {journal} {Phys. Rev. B}\ }\textbf {\bibinfo {volume} {85}},\
  \bibinfo {pages} {195130} (\bibinfo {year} {2012})}\BibitemShut {NoStop}%
\bibitem [{\citenamefont {Karcher}\ \emph {et~al.}(2022)\citenamefont
  {Karcher}, \citenamefont {Gruzberg},\ and\ \citenamefont
  {Mirlin}}]{PhysRevB.106.104202}%
  \BibitemOpen
  \bibfield  {author} {\bibinfo {author} {\bibfnamefont {J.~F.}\ \bibnamefont
  {Karcher}}, \bibinfo {author} {\bibfnamefont {I.~A.}\ \bibnamefont
  {Gruzberg}},\ and\ \bibinfo {author} {\bibfnamefont {A.~D.}\ \bibnamefont
  {Mirlin}},\ }\bibfield  {title} {\bibinfo {title} {Generalized
  multifractality at metal-insulator transitions and in metallic phases of
  two-dimensional disordered systems},\ }\href
  {https://doi.org/10.1103/PhysRevB.106.104202} {\bibfield  {journal} {\bibinfo
   {journal} {Phys. Rev. B}\ }\textbf {\bibinfo {volume} {106}},\ \bibinfo
  {pages} {104202} (\bibinfo {year} {2022})}\BibitemShut {NoStop}%
\bibitem [{\citenamefont {Nomura}\ \emph {et~al.}(2007)\citenamefont {Nomura},
  \citenamefont {Koshino},\ and\ \citenamefont
  {Ryu}}]{nomura_topological_2007}%
  \BibitemOpen
  \bibfield  {author} {\bibinfo {author} {\bibfnamefont {K.}~\bibnamefont
  {Nomura}}, \bibinfo {author} {\bibfnamefont {M.}~\bibnamefont {Koshino}},\
  and\ \bibinfo {author} {\bibfnamefont {S.}~\bibnamefont {Ryu}},\ }\bibfield
  {title} {\bibinfo {title} {Topological {Delocalization} of
  {Two}-{Dimensional} {Massless} {Dirac} {Fermions}},\ }\href
  {https://doi.org/10.1103/PhysRevLett.99.146806} {\bibfield  {journal}
  {\bibinfo  {journal} {Physical Review Letters}\ }\textbf {\bibinfo {volume}
  {99}},\ \bibinfo {pages} {146806} (\bibinfo {year} {2007})},\ \bibinfo {note}
  {publisher: American Physical Society}\BibitemShut {NoStop}%
\bibitem [{\citenamefont {Bardarson}\ \emph {et~al.}(2007)\citenamefont
  {Bardarson}, \citenamefont {Tworzyd\l{}o}, \citenamefont {Brouwer},\ and\
  \citenamefont {Beenakker}}]{bardarson_one-parameter_2007}%
  \BibitemOpen
  \bibfield  {author} {\bibinfo {author} {\bibfnamefont {J.~H.}\ \bibnamefont
  {Bardarson}}, \bibinfo {author} {\bibfnamefont {J.}~\bibnamefont
  {Tworzyd\l{}o}}, \bibinfo {author} {\bibfnamefont {P.~W.}\ \bibnamefont
  {Brouwer}},\ and\ \bibinfo {author} {\bibfnamefont {C.~W.~J.}\ \bibnamefont
  {Beenakker}},\ }\bibfield  {title} {\bibinfo {title} {One-{Parameter}
  {Scaling} at the {Dirac} {Point} in {Graphene}},\ }\href
  {https://doi.org/10.1103/PhysRevLett.99.106801} {\bibfield  {journal}
  {\bibinfo  {journal} {Physical Review Letters}\ }\textbf {\bibinfo {volume}
  {99}},\ \bibinfo {pages} {106801} (\bibinfo {year} {2007})},\ \bibinfo {note}
  {publisher: American Physical Society}\BibitemShut {NoStop}%
\bibitem [{\citenamefont {Tworzyd\l{}o}\ \emph {et~al.}(2008)\citenamefont
  {Tworzyd\l{}o}, \citenamefont {Groth},\ and\ \citenamefont
  {Beenakker}}]{tworzydlo_finite_2008}%
  \BibitemOpen
  \bibfield  {author} {\bibinfo {author} {\bibfnamefont {J.}~\bibnamefont
  {Tworzyd\l{}o}}, \bibinfo {author} {\bibfnamefont {C.~W.}\ \bibnamefont
  {Groth}},\ and\ \bibinfo {author} {\bibfnamefont {C.~W.~J.}\ \bibnamefont
  {Beenakker}},\ }\bibfield  {title} {\bibinfo {title} {Finite difference
  method for transport properties of massless {Dirac} fermions},\ }\href
  {https://doi.org/10.1103/PhysRevB.78.235438} {\bibfield  {journal} {\bibinfo
  {journal} {Physical Review B}\ }\textbf {\bibinfo {volume} {78}},\ \bibinfo
  {pages} {235438} (\bibinfo {year} {2008})},\ \bibinfo {note} {publisher:
  American Physical Society}\BibitemShut {NoStop}%
\bibitem [{\citenamefont {Bardarson}\ \emph {et~al.}(2010)\citenamefont
  {Bardarson}, \citenamefont {Medvedyeva}, \citenamefont {Tworzyd\l{}o},
  \citenamefont {Akhmerov},\ and\ \citenamefont
  {Beenakker}}]{bardarson_absence_2010}%
  \BibitemOpen
  \bibfield  {author} {\bibinfo {author} {\bibfnamefont {J.~H.}\ \bibnamefont
  {Bardarson}}, \bibinfo {author} {\bibfnamefont {M.~V.}\ \bibnamefont
  {Medvedyeva}}, \bibinfo {author} {\bibfnamefont {J.}~\bibnamefont
  {Tworzyd\l{}o}}, \bibinfo {author} {\bibfnamefont {A.~R.}\ \bibnamefont
  {Akhmerov}},\ and\ \bibinfo {author} {\bibfnamefont {C.~W.~J.}\ \bibnamefont
  {Beenakker}},\ }\bibfield  {title} {\bibinfo {title} {Absence of a metallic
  phase in charge-neutral graphene with a random gap},\ }\href
  {https://doi.org/10.1103/PhysRevB.81.121414} {\bibfield  {journal} {\bibinfo
  {journal} {Physical Review B}\ }\textbf {\bibinfo {volume} {81}},\ \bibinfo
  {pages} {121414} (\bibinfo {year} {2010})},\ \bibinfo {note} {publisher:
  American Physical Society}\BibitemShut {NoStop}%
\bibitem [{\citenamefont {Medvedyeva}\ \emph {et~al.}(2010)\citenamefont
  {Medvedyeva}, \citenamefont {Tworzyd\l{}o},\ and\ \citenamefont
  {Beenakker}}]{medvedyeva_effective_2010}%
  \BibitemOpen
  \bibfield  {author} {\bibinfo {author} {\bibfnamefont {M.~V.}\ \bibnamefont
  {Medvedyeva}}, \bibinfo {author} {\bibfnamefont {J.}~\bibnamefont
  {Tworzyd\l{}o}},\ and\ \bibinfo {author} {\bibfnamefont {C.~W.~J.}\
  \bibnamefont {Beenakker}},\ }\bibfield  {title} {\bibinfo {title} {Effective
  mass and tricritical point for lattice fermions localized by a random mass},\
  }\href {https://doi.org/10.1103/PhysRevB.81.214203} {\bibfield  {journal}
  {\bibinfo  {journal} {Physical Review B}\ }\textbf {\bibinfo {volume} {81}},\
  \bibinfo {pages} {214203} (\bibinfo {year} {2010})},\ \bibinfo {note}
  {publisher: American Physical Society}\BibitemShut {NoStop}%
\bibitem [{\citenamefont {Medvedyeva}(2011)}]{medvedyeva_localization_2011}%
  \BibitemOpen
  \bibfield  {author} {\bibinfo {author} {\bibfnamefont {M.~V.}\ \bibnamefont
  {Medvedyeva}},\ }\emph {\bibinfo {title} {On localization of {Dirac} fermions
  by disorder}},\ \href {https://hdl.handle.net/1887/17606} {Ph.D. thesis},\
  \bibinfo  {school} {Leiden University} (\bibinfo {year} {2011}),\ \bibinfo
  {note} {iSBN: 9789085930990}\BibitemShut {NoStop}%
\bibitem [{\citenamefont {Schuessler}\ \emph {et~al.}(2009)\citenamefont
  {Schuessler}, \citenamefont {Ostrovsky}, \citenamefont {Gornyi},\ and\
  \citenamefont {Mirlin}}]{PhysRevB.79.075405}%
  \BibitemOpen
  \bibfield  {author} {\bibinfo {author} {\bibfnamefont {A.}~\bibnamefont
  {Schuessler}}, \bibinfo {author} {\bibfnamefont {P.~M.}\ \bibnamefont
  {Ostrovsky}}, \bibinfo {author} {\bibfnamefont {I.~V.}\ \bibnamefont
  {Gornyi}},\ and\ \bibinfo {author} {\bibfnamefont {A.~D.}\ \bibnamefont
  {Mirlin}},\ }\bibfield  {title} {\bibinfo {title} {Analytic theory of
  ballistic transport in disordered graphene},\ }\href
  {https://doi.org/10.1103/PhysRevB.79.075405} {\bibfield  {journal} {\bibinfo
  {journal} {Phys. Rev. B}\ }\textbf {\bibinfo {volume} {79}},\ \bibinfo
  {pages} {075405} (\bibinfo {year} {2009})}\BibitemShut {NoStop}%
\bibitem [{\citenamefont {Lian}\ \emph {et~al.}(2018)\citenamefont {Lian},
  \citenamefont {Wang}, \citenamefont {Sun}, \citenamefont {Vaezi},\ and\
  \citenamefont {Zhang}}]{PhysRevB.97.125408}%
  \BibitemOpen
  \bibfield  {author} {\bibinfo {author} {\bibfnamefont {B.}~\bibnamefont
  {Lian}}, \bibinfo {author} {\bibfnamefont {J.}~\bibnamefont {Wang}}, \bibinfo
  {author} {\bibfnamefont {X.-Q.}\ \bibnamefont {Sun}}, \bibinfo {author}
  {\bibfnamefont {A.}~\bibnamefont {Vaezi}},\ and\ \bibinfo {author}
  {\bibfnamefont {S.-C.}\ \bibnamefont {Zhang}},\ }\bibfield  {title} {\bibinfo
  {title} {Quantum phase transition of chiral majorana fermions in the presence
  of disorder},\ }\href {https://doi.org/10.1103/PhysRevB.97.125408} {\bibfield
   {journal} {\bibinfo  {journal} {Phys. Rev. B}\ }\textbf {\bibinfo {volume}
  {97}},\ \bibinfo {pages} {125408} (\bibinfo {year} {2018})}\BibitemShut
  {NoStop}%
\bibitem [{\citenamefont {Luo}\ \emph {et~al.}(2022)\citenamefont {Luo},
  \citenamefont {Xiao}, \citenamefont {Kawabata}, \citenamefont {Ohtsuki},\
  and\ \citenamefont {Shindou}}]{PhysRevResearch.4.L022035}%
  \BibitemOpen
  \bibfield  {author} {\bibinfo {author} {\bibfnamefont {X.}~\bibnamefont
  {Luo}}, \bibinfo {author} {\bibfnamefont {Z.}~\bibnamefont {Xiao}}, \bibinfo
  {author} {\bibfnamefont {K.}~\bibnamefont {Kawabata}}, \bibinfo {author}
  {\bibfnamefont {T.}~\bibnamefont {Ohtsuki}},\ and\ \bibinfo {author}
  {\bibfnamefont {R.}~\bibnamefont {Shindou}},\ }\bibfield  {title} {\bibinfo
  {title} {Unifying the anderson transitions in hermitian and non-hermitian
  systems},\ }\href {https://doi.org/10.1103/PhysRevResearch.4.L022035}
  {\bibfield  {journal} {\bibinfo  {journal} {Phys. Rev. Res.}\ }\textbf
  {\bibinfo {volume} {4}},\ \bibinfo {pages} {L022035} (\bibinfo {year}
  {2022})}\BibitemShut {NoStop}%
\bibitem [{\citenamefont {Geim}\ and\ \citenamefont
  {Novoselov}(2007)}]{geim2007rise}%
  \BibitemOpen
  \bibfield  {author} {\bibinfo {author} {\bibfnamefont {A.~K.}\ \bibnamefont
  {Geim}}\ and\ \bibinfo {author} {\bibfnamefont {K.~S.}\ \bibnamefont
  {Novoselov}},\ }\bibfield  {title} {\bibinfo {title} {The rise of graphene},\
  }\href {https://www.nature.com/articles/nmat1849} {\bibfield  {journal}
  {\bibinfo  {journal} {Nature materials}\ }\textbf {\bibinfo {volume} {6}},\
  \bibinfo {pages} {183} (\bibinfo {year} {2007})}\BibitemShut {NoStop}%
\bibitem [{\citenamefont {Kallin}\ and\ \citenamefont
  {Berlinsky}(2009)}]{kallin2009sr2ruo4}%
  \BibitemOpen
  \bibfield  {author} {\bibinfo {author} {\bibfnamefont {C.}~\bibnamefont
  {Kallin}}\ and\ \bibinfo {author} {\bibfnamefont {A.}~\bibnamefont
  {Berlinsky}},\ }\bibfield  {title} {\bibinfo {title} {Is sr2ruo4 a chiral
  p-wave superconductor?},\ }\href
  {https://iopscience.iop.org/article/10.1088/0953-8984/21/16/164210/meta}
  {\bibfield  {journal} {\bibinfo  {journal} {Journal of Physics: Condensed
  Matter}\ }\textbf {\bibinfo {volume} {21}},\ \bibinfo {pages} {164210}
  (\bibinfo {year} {2009})}\BibitemShut {NoStop}%
\bibitem [{\citenamefont {Tewari}\ \emph {et~al.}(2007)\citenamefont {Tewari},
  \citenamefont {Das~Sarma}, \citenamefont {Nayak}, \citenamefont {Zhang},\
  and\ \citenamefont {Zoller}}]{PhysRevLett.98.010506}%
  \BibitemOpen
  \bibfield  {author} {\bibinfo {author} {\bibfnamefont {S.}~\bibnamefont
  {Tewari}}, \bibinfo {author} {\bibfnamefont {S.}~\bibnamefont {Das~Sarma}},
  \bibinfo {author} {\bibfnamefont {C.}~\bibnamefont {Nayak}}, \bibinfo
  {author} {\bibfnamefont {C.}~\bibnamefont {Zhang}},\ and\ \bibinfo {author}
  {\bibfnamefont {P.}~\bibnamefont {Zoller}},\ }\bibfield  {title} {\bibinfo
  {title} {Quantum computation using vortices and majorana zero modes of a
  ${p}_{x}+i{p}_{y}$ superfluid of fermionic cold atoms},\ }\href
  {https://doi.org/10.1103/PhysRevLett.98.010506} {\bibfield  {journal}
  {\bibinfo  {journal} {Phys. Rev. Lett.}\ }\textbf {\bibinfo {volume} {98}},\
  \bibinfo {pages} {010506} (\bibinfo {year} {2007})}\BibitemShut {NoStop}%
\bibitem [{\citenamefont {Chen}\ \emph {et~al.}(2023)\citenamefont {Chen},
  \citenamefont {Fu}, \citenamefont {Xu}, \citenamefont {Shi}, \citenamefont
  {Cui},\ and\ \citenamefont {Zhang}}]{PhysRevB.108.064208}%
  \BibitemOpen
  \bibfield  {author} {\bibinfo {author} {\bibfnamefont {Y.}~\bibnamefont
  {Chen}}, \bibinfo {author} {\bibfnamefont {B.}~\bibnamefont {Fu}}, \bibinfo
  {author} {\bibfnamefont {J.}~\bibnamefont {Xu}}, \bibinfo {author}
  {\bibfnamefont {Q.}~\bibnamefont {Shi}}, \bibinfo {author} {\bibfnamefont
  {P.}~\bibnamefont {Cui}},\ and\ \bibinfo {author} {\bibfnamefont
  {Z.}~\bibnamefont {Zhang}},\ }\bibfield  {title} {\bibinfo {title}
  {Quasiparticle and transport properties of disordered bilayer graphene},\
  }\href {https://doi.org/10.1103/PhysRevB.108.064208} {\bibfield  {journal}
  {\bibinfo  {journal} {Phys. Rev. B}\ }\textbf {\bibinfo {volume} {108}},\
  \bibinfo {pages} {064208} (\bibinfo {year} {2023})}\BibitemShut {NoStop}%
\bibitem [{\citenamefont {Fu}\ \emph {et~al.}(2023)\citenamefont {Fu},
  \citenamefont {Chen}, \citenamefont {Chen}, \citenamefont {Zhu},
  \citenamefont {Cui}, \citenamefont {Li}, \citenamefont {Zhang},\ and\
  \citenamefont {Shi}}]{PhysRevB.108.064207}%
  \BibitemOpen
  \bibfield  {author} {\bibinfo {author} {\bibfnamefont {B.}~\bibnamefont
  {Fu}}, \bibinfo {author} {\bibfnamefont {Y.}~\bibnamefont {Chen}}, \bibinfo
  {author} {\bibfnamefont {W.}~\bibnamefont {Chen}}, \bibinfo {author}
  {\bibfnamefont {W.}~\bibnamefont {Zhu}}, \bibinfo {author} {\bibfnamefont
  {P.}~\bibnamefont {Cui}}, \bibinfo {author} {\bibfnamefont {Q.}~\bibnamefont
  {Li}}, \bibinfo {author} {\bibfnamefont {Z.}~\bibnamefont {Zhang}},\ and\
  \bibinfo {author} {\bibfnamefont {Q.}~\bibnamefont {Shi}},\ }\bibfield
  {title} {\bibinfo {title} {Disorder effects on the quasiparticle and
  transport properties of two-dimensional dirac fermionic systems},\ }\href
  {https://doi.org/10.1103/PhysRevB.108.064207} {\bibfield  {journal} {\bibinfo
   {journal} {Phys. Rev. B}\ }\textbf {\bibinfo {volume} {108}},\ \bibinfo
  {pages} {064207} (\bibinfo {year} {2023})}\BibitemShut {NoStop}%
\bibitem [{\citenamefont {Anwar}\ \emph {et~al.}(2020)\citenamefont {Anwar},
  \citenamefont {Iurov}, \citenamefont {Huang}, \citenamefont {Gumbs},\ and\
  \citenamefont {Sharma}}]{PhysRevB.101.115424}%
  \BibitemOpen
  \bibfield  {author} {\bibinfo {author} {\bibfnamefont {F.}~\bibnamefont
  {Anwar}}, \bibinfo {author} {\bibfnamefont {A.}~\bibnamefont {Iurov}},
  \bibinfo {author} {\bibfnamefont {D.}~\bibnamefont {Huang}}, \bibinfo
  {author} {\bibfnamefont {G.}~\bibnamefont {Gumbs}},\ and\ \bibinfo {author}
  {\bibfnamefont {A.}~\bibnamefont {Sharma}},\ }\bibfield  {title} {\bibinfo
  {title} {Interplay between effects of barrier tilting and scatterers within a
  barrier on tunneling transport of dirac electrons in graphene},\ }\href
  {https://doi.org/10.1103/PhysRevB.101.115424} {\bibfield  {journal} {\bibinfo
   {journal} {Phys. Rev. B}\ }\textbf {\bibinfo {volume} {101}},\ \bibinfo
  {pages} {115424} (\bibinfo {year} {2020})}\BibitemShut {NoStop}%
\bibitem [{\citenamefont {Couto}\ \emph {et~al.}(2014)\citenamefont {Couto},
  \citenamefont {Costanzo}, \citenamefont {Engels}, \citenamefont {Ki},
  \citenamefont {Watanabe}, \citenamefont {Taniguchi}, \citenamefont
  {Stampfer}, \citenamefont {Guinea},\ and\ \citenamefont
  {Morpurgo}}]{PhysRevX.4.041019}%
  \BibitemOpen
  \bibfield  {author} {\bibinfo {author} {\bibfnamefont {N.~J.~G.}\
  \bibnamefont {Couto}}, \bibinfo {author} {\bibfnamefont {D.}~\bibnamefont
  {Costanzo}}, \bibinfo {author} {\bibfnamefont {S.}~\bibnamefont {Engels}},
  \bibinfo {author} {\bibfnamefont {D.-K.}\ \bibnamefont {Ki}}, \bibinfo
  {author} {\bibfnamefont {K.}~\bibnamefont {Watanabe}}, \bibinfo {author}
  {\bibfnamefont {T.}~\bibnamefont {Taniguchi}}, \bibinfo {author}
  {\bibfnamefont {C.}~\bibnamefont {Stampfer}}, \bibinfo {author}
  {\bibfnamefont {F.}~\bibnamefont {Guinea}},\ and\ \bibinfo {author}
  {\bibfnamefont {A.~F.}\ \bibnamefont {Morpurgo}},\ }\bibfield  {title}
  {\bibinfo {title} {Random strain fluctuations as dominant disorder source for
  high-quality on-substrate graphene devices},\ }\href
  {https://doi.org/10.1103/PhysRevX.4.041019} {\bibfield  {journal} {\bibinfo
  {journal} {Phys. Rev. X}\ }\textbf {\bibinfo {volume} {4}},\ \bibinfo {pages}
  {041019} (\bibinfo {year} {2014})}\BibitemShut {NoStop}%
\bibitem [{\citenamefont {Meng}\ \emph {et~al.}(2021)\citenamefont {Meng},
  \citenamefont {Mondaini}, \citenamefont {Ma},\ and\ \citenamefont
  {Lin}}]{PhysRevB.104.045138}%
  \BibitemOpen
  \bibfield  {author} {\bibinfo {author} {\bibfnamefont {J.}~\bibnamefont
  {Meng}}, \bibinfo {author} {\bibfnamefont {R.}~\bibnamefont {Mondaini}},
  \bibinfo {author} {\bibfnamefont {T.}~\bibnamefont {Ma}},\ and\ \bibinfo
  {author} {\bibfnamefont {H.-Q.}\ \bibnamefont {Lin}},\ }\bibfield  {title}
  {\bibinfo {title} {Inducing a metal-insulator transition in disordered
  interacting dirac fermion systems via an external magnetic field},\ }\href
  {https://doi.org/10.1103/PhysRevB.104.045138} {\bibfield  {journal} {\bibinfo
   {journal} {Phys. Rev. B}\ }\textbf {\bibinfo {volume} {104}},\ \bibinfo
  {pages} {045138} (\bibinfo {year} {2021})}\BibitemShut {NoStop}%
\bibitem [{\citenamefont {Hui}\ \emph {et~al.}(2020)\citenamefont {Hui},
  \citenamefont {Lederer}, \citenamefont {Oganesyan},\ and\ \citenamefont
  {Kim}}]{PhysRevB.101.121107}%
  \BibitemOpen
  \bibfield  {author} {\bibinfo {author} {\bibfnamefont {A.}~\bibnamefont
  {Hui}}, \bibinfo {author} {\bibfnamefont {S.}~\bibnamefont {Lederer}},
  \bibinfo {author} {\bibfnamefont {V.}~\bibnamefont {Oganesyan}},\ and\
  \bibinfo {author} {\bibfnamefont {E.-A.}\ \bibnamefont {Kim}},\ }\bibfield
  {title} {\bibinfo {title} {Quantum aspects of hydrodynamic transport from
  weak electron-impurity scattering},\ }\href
  {https://doi.org/10.1103/PhysRevB.101.121107} {\bibfield  {journal} {\bibinfo
   {journal} {Phys. Rev. B}\ }\textbf {\bibinfo {volume} {101}},\ \bibinfo
  {pages} {121107} (\bibinfo {year} {2020})}\BibitemShut {NoStop}%
\bibitem [{\citenamefont {Ostrovsky}\ \emph {et~al.}(2007)\citenamefont
  {Ostrovsky}, \citenamefont {Gornyi},\ and\ \citenamefont
  {Mirlin}}]{ostrovsky_quantum_2007}%
  \BibitemOpen
  \bibfield  {author} {\bibinfo {author} {\bibfnamefont {P.~M.}\ \bibnamefont
  {Ostrovsky}}, \bibinfo {author} {\bibfnamefont {I.~V.}\ \bibnamefont
  {Gornyi}},\ and\ \bibinfo {author} {\bibfnamefont {A.~D.}\ \bibnamefont
  {Mirlin}},\ }\bibfield  {title} {\bibinfo {title} {Quantum {Criticality} and
  {Minimal} {Conductivity} in {Graphene} with {Long}-{Range} {Disorder}},\
  }\href {https://doi.org/10.1103/PhysRevLett.98.256801} {\bibfield  {journal}
  {\bibinfo  {journal} {Physical Review Letters}\ }\textbf {\bibinfo {volume}
  {98}},\ \bibinfo {pages} {256801} (\bibinfo {year} {2007})},\ \bibinfo {note}
  {publisher: American Physical Society}\BibitemShut {NoStop}%
\bibitem [{\citenamefont {Tworzyd\l{}o}\ \emph {et~al.}(2006)\citenamefont
  {Tworzyd\l{}o}, \citenamefont {Trauzettel}, \citenamefont {Titov},
  \citenamefont {Rycerz},\ and\ \citenamefont
  {Beenakker}}]{tworzydlo_sub-poissonian_2006}%
  \BibitemOpen
  \bibfield  {author} {\bibinfo {author} {\bibfnamefont {J.}~\bibnamefont
  {Tworzyd\l{}o}}, \bibinfo {author} {\bibfnamefont {B.}~\bibnamefont
  {Trauzettel}}, \bibinfo {author} {\bibfnamefont {M.}~\bibnamefont {Titov}},
  \bibinfo {author} {\bibfnamefont {A.}~\bibnamefont {Rycerz}},\ and\ \bibinfo
  {author} {\bibfnamefont {C.~W.~J.}\ \bibnamefont {Beenakker}},\ }\bibfield
  {title} {\bibinfo {title} {Sub-{Poissonian} {Shot} {Noise} in {Graphene}},\
  }\href {https://doi.org/10.1103/PhysRevLett.96.246802} {\bibfield  {journal}
  {\bibinfo  {journal} {Physical Review Letters}\ }\textbf {\bibinfo {volume}
  {96}},\ \bibinfo {pages} {246802} (\bibinfo {year} {2006})},\ \bibinfo {note}
  {publisher: American Physical Society}\BibitemShut {NoStop}%
\bibitem [{\citenamefont {Katsnelson}(2006)}]{katsnelson2006zitterbewegung}%
  \BibitemOpen
  \bibfield  {author} {\bibinfo {author} {\bibfnamefont {M.}~\bibnamefont
  {Katsnelson}},\ }\bibfield  {title} {\bibinfo {title} {Zitterbewegung,
  chirality, and minimal conductivity in graphene},\ }\href
  {https://link.springer.com/article/10.1140/epjb/e2006-00203-1} {\bibfield
  {journal} {\bibinfo  {journal} {The European Physical Journal B-Condensed
  Matter and Complex Systems}\ }\textbf {\bibinfo {volume} {51}},\ \bibinfo
  {pages} {157} (\bibinfo {year} {2006})}\BibitemShut {NoStop}%
\bibitem [{\citenamefont {Wang}\ \emph {et~al.}(2021)\citenamefont {Wang},
  \citenamefont {Pan}, \citenamefont {Ohtsuki}, \citenamefont {Gruzberg},\ and\
  \citenamefont {Shindou}}]{PhysRevB.104.184201}%
  \BibitemOpen
  \bibfield  {author} {\bibinfo {author} {\bibfnamefont {T.}~\bibnamefont
  {Wang}}, \bibinfo {author} {\bibfnamefont {Z.}~\bibnamefont {Pan}}, \bibinfo
  {author} {\bibfnamefont {T.}~\bibnamefont {Ohtsuki}}, \bibinfo {author}
  {\bibfnamefont {I.~A.}\ \bibnamefont {Gruzberg}},\ and\ \bibinfo {author}
  {\bibfnamefont {R.}~\bibnamefont {Shindou}},\ }\bibfield  {title} {\bibinfo
  {title} {Multicriticality of two-dimensional class-d disordered topological
  superconductors},\ }\href {https://doi.org/10.1103/PhysRevB.104.184201}
  {\bibfield  {journal} {\bibinfo  {journal} {Phys. Rev. B}\ }\textbf {\bibinfo
  {volume} {104}},\ \bibinfo {pages} {184201} (\bibinfo {year}
  {2021})}\BibitemShut {NoStop}%
\bibitem [{\citenamefont {Pan}\ \emph {et~al.}(2021)\citenamefont {Pan},
  \citenamefont {Wang}, \citenamefont {Ohtsuki},\ and\ \citenamefont
  {Shindou}}]{PhysRevB.104.174205}%
  \BibitemOpen
  \bibfield  {author} {\bibinfo {author} {\bibfnamefont {Z.}~\bibnamefont
  {Pan}}, \bibinfo {author} {\bibfnamefont {T.}~\bibnamefont {Wang}}, \bibinfo
  {author} {\bibfnamefont {T.}~\bibnamefont {Ohtsuki}},\ and\ \bibinfo {author}
  {\bibfnamefont {R.}~\bibnamefont {Shindou}},\ }\bibfield  {title} {\bibinfo
  {title} {Renormalization group analysis of dirac fermions with a random
  mass},\ }\href {https://doi.org/10.1103/PhysRevB.104.174205} {\bibfield
  {journal} {\bibinfo  {journal} {Phys. Rev. B}\ }\textbf {\bibinfo {volume}
  {104}},\ \bibinfo {pages} {174205} (\bibinfo {year} {2021})}\BibitemShut
  {NoStop}%
\bibitem [{\citenamefont {Titov}\ and\ \citenamefont
  {Katsnelson}(2014)}]{titov_metal-insulator_2014}%
  \BibitemOpen
  \bibfield  {author} {\bibinfo {author} {\bibfnamefont {M.}~\bibnamefont
  {Titov}}\ and\ \bibinfo {author} {\bibfnamefont {M.}~\bibnamefont
  {Katsnelson}},\ }\bibfield  {title} {\bibinfo {title} {Metal-{Insulator}
  {Transition} in {Graphene} on {Boron} {Nitride}},\ }\href
  {https://doi.org/10.1103/PhysRevLett.113.096801} {\bibfield  {journal}
  {\bibinfo  {journal} {Physical Review Letters}\ }\textbf {\bibinfo {volume}
  {113}},\ \bibinfo {pages} {096801} (\bibinfo {year} {2014})},\ \bibinfo
  {note} {publisher: American Physical Society}\BibitemShut {NoStop}%
\bibitem [{\citenamefont {Altland}\ \emph {et~al.}(2002)\citenamefont
  {Altland}, \citenamefont {Simons},\ and\ \citenamefont
  {Zirnbauer}}]{altland2002theories}%
  \BibitemOpen
  \bibfield  {author} {\bibinfo {author} {\bibfnamefont {A.}~\bibnamefont
  {Altland}}, \bibinfo {author} {\bibfnamefont {B.}~\bibnamefont {Simons}},\
  and\ \bibinfo {author} {\bibfnamefont {M.}~\bibnamefont {Zirnbauer}},\
  }\bibfield  {title} {\bibinfo {title} {Theories of low-energy quasi-particle
  states in disordered d-wave superconductors},\ }\href
  {https://doi.org/10.1016/S0370-1573(01)00065-5} {\bibfield  {journal}
  {\bibinfo  {journal} {Physics Reports}\ }\textbf {\bibinfo {volume} {359}},\
  \bibinfo {pages} {283} (\bibinfo {year} {2002})}\BibitemShut {NoStop}%
\bibitem [{\citenamefont {Nomura}\ \emph {et~al.}(2008)\citenamefont {Nomura},
  \citenamefont {Ryu}, \citenamefont {Koshino}, \citenamefont {Mudry},\ and\
  \citenamefont {Furusaki}}]{nomura_quantum_2008}%
  \BibitemOpen
  \bibfield  {author} {\bibinfo {author} {\bibfnamefont {K.}~\bibnamefont
  {Nomura}}, \bibinfo {author} {\bibfnamefont {S.}~\bibnamefont {Ryu}},
  \bibinfo {author} {\bibfnamefont {M.}~\bibnamefont {Koshino}}, \bibinfo
  {author} {\bibfnamefont {C.}~\bibnamefont {Mudry}},\ and\ \bibinfo {author}
  {\bibfnamefont {A.}~\bibnamefont {Furusaki}},\ }\bibfield  {title} {\bibinfo
  {title} {Quantum {Hall} {Effect} of {Massless} {Dirac} {Fermions} in a
  {Vanishing} {Magnetic} {Field}},\ }\href
  {https://doi.org/10.1103/PhysRevLett.100.246806} {\bibfield  {journal}
  {\bibinfo  {journal} {Physical Review Letters}\ }\textbf {\bibinfo {volume}
  {100}},\ \bibinfo {pages} {246806} (\bibinfo {year} {2008})},\ \bibinfo
  {note} {publisher: American Physical Society}\BibitemShut {NoStop}%
\bibitem [{\citenamefont {Hill}\ and\ \citenamefont
  {Ziegler}(2014)}]{hill2014scaling}%
  \BibitemOpen
  \bibfield  {author} {\bibinfo {author} {\bibfnamefont {A.}~\bibnamefont
  {Hill}}\ and\ \bibinfo {author} {\bibfnamefont {K.}~\bibnamefont {Ziegler}},\
  }\bibfield  {title} {\bibinfo {title} {Scaling behavior of disordered lattice
  fermions in two dimensions},\ }\href
  {https://link.springer.com/article/10.1140/epjb/e2014-41073-x} {\bibfield
  {journal} {\bibinfo  {journal} {The European Physical Journal B}\ }\textbf
  {\bibinfo {volume} {87}},\ \bibinfo {pages} {1} (\bibinfo {year}
  {2014})}\BibitemShut {NoStop}%
\bibitem [{\citenamefont {Sbierski}\ \emph {et~al.}(2021)\citenamefont
  {Sbierski}, \citenamefont {Dresselhaus}, \citenamefont {Moore},\ and\
  \citenamefont {Gruzberg}}]{PhysRevLett.126.076801}%
  \BibitemOpen
  \bibfield  {author} {\bibinfo {author} {\bibfnamefont {B.}~\bibnamefont
  {Sbierski}}, \bibinfo {author} {\bibfnamefont {E.~J.}\ \bibnamefont
  {Dresselhaus}}, \bibinfo {author} {\bibfnamefont {J.~E.}\ \bibnamefont
  {Moore}},\ and\ \bibinfo {author} {\bibfnamefont {I.~A.}\ \bibnamefont
  {Gruzberg}},\ }\bibfield  {title} {\bibinfo {title} {Criticality of
  two-dimensional disordered dirac fermions in the unitary class and
  universality of the integer quantum hall transition},\ }\href
  {https://doi.org/10.1103/PhysRevLett.126.076801} {\bibfield  {journal}
  {\bibinfo  {journal} {Phys. Rev. Lett.}\ }\textbf {\bibinfo {volume} {126}},\
  \bibinfo {pages} {076801} (\bibinfo {year} {2021})}\BibitemShut {NoStop}%
\bibitem [{\citenamefont {Decker}\ \emph {et~al.}(2011)\citenamefont {Decker},
  \citenamefont {Wang}, \citenamefont {Brar}, \citenamefont {Regan},
  \citenamefont {Tsai}, \citenamefont {Wu}, \citenamefont {Gannett},
  \citenamefont {Zettl},\ and\ \citenamefont {Crommie}}]{decker2011local}%
  \BibitemOpen
  \bibfield  {author} {\bibinfo {author} {\bibfnamefont {R.}~\bibnamefont
  {Decker}}, \bibinfo {author} {\bibfnamefont {Y.}~\bibnamefont {Wang}},
  \bibinfo {author} {\bibfnamefont {V.~W.}\ \bibnamefont {Brar}}, \bibinfo
  {author} {\bibfnamefont {W.}~\bibnamefont {Regan}}, \bibinfo {author}
  {\bibfnamefont {H.-Z.}\ \bibnamefont {Tsai}}, \bibinfo {author}
  {\bibfnamefont {Q.}~\bibnamefont {Wu}}, \bibinfo {author} {\bibfnamefont
  {W.}~\bibnamefont {Gannett}}, \bibinfo {author} {\bibfnamefont
  {A.}~\bibnamefont {Zettl}},\ and\ \bibinfo {author} {\bibfnamefont {M.~F.}\
  \bibnamefont {Crommie}},\ }\bibfield  {title} {\bibinfo {title} {Local
  electronic properties of graphene on a bn substrate via scanning tunneling
  microscopy},\ }\href {https://pubs.acs.org/doi/full/10.1021/nl2005115}
  {\bibfield  {journal} {\bibinfo  {journal} {Nano letters}\ }\textbf {\bibinfo
  {volume} {11}},\ \bibinfo {pages} {2291} (\bibinfo {year}
  {2011})}\BibitemShut {NoStop}%
\bibitem [{\citenamefont {Dean}\ \emph {et~al.}(2010)\citenamefont {Dean},
  \citenamefont {Young}, \citenamefont {Meric}, \citenamefont {Lee},
  \citenamefont {Wang}, \citenamefont {Sorgenfrei}, \citenamefont {Watanabe},
  \citenamefont {Taniguchi}, \citenamefont {Kim}, \citenamefont {Shepard} \emph
  {et~al.}}]{dean2010boron}%
  \BibitemOpen
  \bibfield  {author} {\bibinfo {author} {\bibfnamefont {C.~R.}\ \bibnamefont
  {Dean}}, \bibinfo {author} {\bibfnamefont {A.~F.}\ \bibnamefont {Young}},
  \bibinfo {author} {\bibfnamefont {I.}~\bibnamefont {Meric}}, \bibinfo
  {author} {\bibfnamefont {C.}~\bibnamefont {Lee}}, \bibinfo {author}
  {\bibfnamefont {L.}~\bibnamefont {Wang}}, \bibinfo {author} {\bibfnamefont
  {S.}~\bibnamefont {Sorgenfrei}}, \bibinfo {author} {\bibfnamefont
  {K.}~\bibnamefont {Watanabe}}, \bibinfo {author} {\bibfnamefont
  {T.}~\bibnamefont {Taniguchi}}, \bibinfo {author} {\bibfnamefont
  {P.}~\bibnamefont {Kim}}, \bibinfo {author} {\bibfnamefont {K.~L.}\
  \bibnamefont {Shepard}}, \emph {et~al.},\ }\bibfield  {title} {\bibinfo
  {title} {Boron nitride substrates for high-quality graphene electronics},\
  }\href {https://www.nature.com/articles/nnano.2010.172.} {\bibfield
  {journal} {\bibinfo  {journal} {Nature nanotechnology}\ }\textbf {\bibinfo
  {volume} {5}},\ \bibinfo {pages} {722} (\bibinfo {year} {2010})}\BibitemShut
  {NoStop}%
\bibitem [{\citenamefont {Xu}\ \emph {et~al.}(2016)\citenamefont {Xu},
  \citenamefont {Song}, \citenamefont {Yuan},\ and\ \citenamefont
  {Zhang}}]{PhysRevB.94.195103}%
  \BibitemOpen
  \bibfield  {author} {\bibinfo {author} {\bibfnamefont {J.-R.}\ \bibnamefont
  {Xu}}, \bibinfo {author} {\bibfnamefont {Z.-Y.}\ \bibnamefont {Song}},
  \bibinfo {author} {\bibfnamefont {C.-G.}\ \bibnamefont {Yuan}},\ and\
  \bibinfo {author} {\bibfnamefont {Y.-Z.}\ \bibnamefont {Zhang}},\ }\bibfield
  {title} {\bibinfo {title} {Interaction-induced metallic state in graphene on
  hexagonal boron nitride},\ }\href
  {https://doi.org/10.1103/PhysRevB.94.195103} {\bibfield  {journal} {\bibinfo
  {journal} {Phys. Rev. B}\ }\textbf {\bibinfo {volume} {94}},\ \bibinfo
  {pages} {195103} (\bibinfo {year} {2016})}\BibitemShut {NoStop}%
\bibitem [{\citenamefont {Li}\ \emph {et~al.}(2020)\citenamefont {Li},
  \citenamefont {Xu},\ and\ \citenamefont {Zhang}}]{li2020band}%
  \BibitemOpen
  \bibfield  {author} {\bibinfo {author} {\bibfnamefont {X.}~\bibnamefont
  {Li}}, \bibinfo {author} {\bibfnamefont {L.}~\bibnamefont {Xu}},\ and\
  \bibinfo {author} {\bibfnamefont {J.}~\bibnamefont {Zhang}},\ }\bibfield
  {title} {\bibinfo {title} {Band structure and transport property of
  graphene/h-bn heterostructure under local potentials},\ }\href
  {https://doi.org/10.1016/j.cjph.2020.02.011} {\bibfield  {journal} {\bibinfo
  {journal} {Chinese Journal of Physics}\ }\textbf {\bibinfo {volume} {65}},\
  \bibinfo {pages} {75} (\bibinfo {year} {2020})}\BibitemShut {NoStop}%
\bibitem [{\citenamefont {Liao}\ \emph {et~al.}(2020)\citenamefont {Liao},
  \citenamefont {Wei}, \citenamefont {Du}, \citenamefont {Wang}, \citenamefont
  {Tang}, \citenamefont {Yu}, \citenamefont {Wu}, \citenamefont {Zhao},
  \citenamefont {Xu}, \citenamefont {Han} \emph {et~al.}}]{liao2020precise}%
  \BibitemOpen
  \bibfield  {author} {\bibinfo {author} {\bibfnamefont {M.}~\bibnamefont
  {Liao}}, \bibinfo {author} {\bibfnamefont {Z.}~\bibnamefont {Wei}}, \bibinfo
  {author} {\bibfnamefont {L.}~\bibnamefont {Du}}, \bibinfo {author}
  {\bibfnamefont {Q.}~\bibnamefont {Wang}}, \bibinfo {author} {\bibfnamefont
  {J.}~\bibnamefont {Tang}}, \bibinfo {author} {\bibfnamefont {H.}~\bibnamefont
  {Yu}}, \bibinfo {author} {\bibfnamefont {F.}~\bibnamefont {Wu}}, \bibinfo
  {author} {\bibfnamefont {J.}~\bibnamefont {Zhao}}, \bibinfo {author}
  {\bibfnamefont {X.}~\bibnamefont {Xu}}, \bibinfo {author} {\bibfnamefont
  {B.}~\bibnamefont {Han}}, \emph {et~al.},\ }\bibfield  {title} {\bibinfo
  {title} {Precise control of the interlayer twist angle in large scale mos2
  homostructures},\ }\href {https://www.nature.com/articles/s41467-020-16056-4}
  {\bibfield  {journal} {\bibinfo  {journal} {Nature communications}\ }\textbf
  {\bibinfo {volume} {11}},\ \bibinfo {pages} {2153} (\bibinfo {year}
  {2020})}\BibitemShut {NoStop}%
\bibitem [{\citenamefont {Sol{\'\i}s-Fern{\'a}ndez}\ \emph
  {et~al.}(2020)\citenamefont {Sol{\'\i}s-Fern{\'a}ndez}, \citenamefont
  {Terao}, \citenamefont {Kawahara}, \citenamefont {Nishiyama}, \citenamefont
  {Uwanno}, \citenamefont {Lin}, \citenamefont {Yamamoto}, \citenamefont
  {Nakashima}, \citenamefont {Nagashio}, \citenamefont {Hibino}, \citenamefont
  {Suenaga},\ and\ \citenamefont {Ago}}]{solis2020isothermal}%
  \BibitemOpen
  \bibfield  {author} {\bibinfo {author} {\bibfnamefont {P.}~\bibnamefont
  {Sol{\'\i}s-Fern{\'a}ndez}}, \bibinfo {author} {\bibfnamefont
  {Y.}~\bibnamefont {Terao}}, \bibinfo {author} {\bibfnamefont
  {K.}~\bibnamefont {Kawahara}}, \bibinfo {author} {\bibfnamefont
  {W.}~\bibnamefont {Nishiyama}}, \bibinfo {author} {\bibfnamefont
  {T.}~\bibnamefont {Uwanno}}, \bibinfo {author} {\bibfnamefont {Y.-C.}\
  \bibnamefont {Lin}}, \bibinfo {author} {\bibfnamefont {K.}~\bibnamefont
  {Yamamoto}}, \bibinfo {author} {\bibfnamefont {H.}~\bibnamefont {Nakashima}},
  \bibinfo {author} {\bibfnamefont {K.}~\bibnamefont {Nagashio}}, \bibinfo
  {author} {\bibfnamefont {H.}~\bibnamefont {Hibino}}, \bibinfo {author}
  {\bibfnamefont {K.}~\bibnamefont {Suenaga}},\ and\ \bibinfo {author}
  {\bibfnamefont {H.}~\bibnamefont {Ago}},\ }\bibfield  {title} {\bibinfo
  {title} {Isothermal growth and stacking evolution in highly uniform
  bernal-stacked bilayer graphene},\ }\href
  {https://doi.org/10.1021/acsnano.0c00645} {\bibfield  {journal} {\bibinfo
  {journal} {ACS Nano}\ }\textbf {\bibinfo {volume} {14}},\ \bibinfo {pages}
  {6834} (\bibinfo {year} {2020})},\ \bibinfo {note} {pMID:
  32407070}\BibitemShut {NoStop}%
\bibitem [{\citenamefont {Brzhezinskaya}\ \emph {et~al.}(2021)\citenamefont
  {Brzhezinskaya}, \citenamefont {Kononenko}, \citenamefont {Matveev},
  \citenamefont {Zotov}, \citenamefont {Khodos}, \citenamefont {Levashov},
  \citenamefont {Volkov}, \citenamefont {Bozhko}, \citenamefont {Chekmazov},\
  and\ \citenamefont {Roshchupkin}}]{brzhezinskaya2021engineering}%
  \BibitemOpen
  \bibfield  {author} {\bibinfo {author} {\bibfnamefont {M.}~\bibnamefont
  {Brzhezinskaya}}, \bibinfo {author} {\bibfnamefont {O.}~\bibnamefont
  {Kononenko}}, \bibinfo {author} {\bibfnamefont {V.}~\bibnamefont {Matveev}},
  \bibinfo {author} {\bibfnamefont {A.}~\bibnamefont {Zotov}}, \bibinfo
  {author} {\bibfnamefont {I.~I.}\ \bibnamefont {Khodos}}, \bibinfo {author}
  {\bibfnamefont {V.}~\bibnamefont {Levashov}}, \bibinfo {author}
  {\bibfnamefont {V.}~\bibnamefont {Volkov}}, \bibinfo {author} {\bibfnamefont
  {S.~I.}\ \bibnamefont {Bozhko}}, \bibinfo {author} {\bibfnamefont {S.~V.}\
  \bibnamefont {Chekmazov}},\ and\ \bibinfo {author} {\bibfnamefont
  {D.}~\bibnamefont {Roshchupkin}},\ }\bibfield  {title} {\bibinfo {title}
  {Engineering of numerous moir{\'e} superlattices in twisted multilayer
  graphene for twistronics and straintronics applications},\ }\href
  {https://pubs.acs.org/doi/full/10.1021/acsnano.1c04286} {\bibfield  {journal}
  {\bibinfo  {journal} {ACS nano}\ }\textbf {\bibinfo {volume} {15}},\ \bibinfo
  {pages} {12358} (\bibinfo {year} {2021})}\BibitemShut {NoStop}%
\bibitem [{\citenamefont {Uri}\ \emph {et~al.}(2023)\citenamefont {Uri},
  \citenamefont {de~la Barrera}, \citenamefont {Randeria}, \citenamefont
  {Rodan-Legrain}, \citenamefont {Devakul}, \citenamefont {Crowley},
  \citenamefont {Paul}, \citenamefont {Watanabe}, \citenamefont {Taniguchi},
  \citenamefont {Lifshitz} \emph {et~al.}}]{uri2023superconductivity}%
  \BibitemOpen
  \bibfield  {author} {\bibinfo {author} {\bibfnamefont {A.}~\bibnamefont
  {Uri}}, \bibinfo {author} {\bibfnamefont {S.~C.}\ \bibnamefont {de~la
  Barrera}}, \bibinfo {author} {\bibfnamefont {M.~T.}\ \bibnamefont
  {Randeria}}, \bibinfo {author} {\bibfnamefont {D.}~\bibnamefont
  {Rodan-Legrain}}, \bibinfo {author} {\bibfnamefont {T.}~\bibnamefont
  {Devakul}}, \bibinfo {author} {\bibfnamefont {P.~J.}\ \bibnamefont
  {Crowley}}, \bibinfo {author} {\bibfnamefont {N.}~\bibnamefont {Paul}},
  \bibinfo {author} {\bibfnamefont {K.}~\bibnamefont {Watanabe}}, \bibinfo
  {author} {\bibfnamefont {T.}~\bibnamefont {Taniguchi}}, \bibinfo {author}
  {\bibfnamefont {R.}~\bibnamefont {Lifshitz}}, \emph {et~al.},\ }\bibfield
  {title} {\bibinfo {title} {Superconductivity and strong interactions in a
  tunable moir{\'e} quasicrystal},\ }\href
  {https://www.nature.com/articles/s41586-023-06294-z} {\bibfield  {journal}
  {\bibinfo  {journal} {Nature}\ }\textbf {\bibinfo {volume} {620}},\ \bibinfo
  {pages} {762} (\bibinfo {year} {2023})}\BibitemShut {NoStop}%
\bibitem [{\citenamefont {Wang}\ \emph {et~al.}(2016)\citenamefont {Wang},
  \citenamefont {Lu}, \citenamefont {Ding}, \citenamefont {Yao}, \citenamefont
  {Yan}, \citenamefont {Wan}, \citenamefont {Deng}, \citenamefont {Wang},
  \citenamefont {Chen}, \citenamefont {Ma} \emph {et~al.}}]{wang2016gaps}%
  \BibitemOpen
  \bibfield  {author} {\bibinfo {author} {\bibfnamefont {E.}~\bibnamefont
  {Wang}}, \bibinfo {author} {\bibfnamefont {X.}~\bibnamefont {Lu}}, \bibinfo
  {author} {\bibfnamefont {S.}~\bibnamefont {Ding}}, \bibinfo {author}
  {\bibfnamefont {W.}~\bibnamefont {Yao}}, \bibinfo {author} {\bibfnamefont
  {M.}~\bibnamefont {Yan}}, \bibinfo {author} {\bibfnamefont {G.}~\bibnamefont
  {Wan}}, \bibinfo {author} {\bibfnamefont {K.}~\bibnamefont {Deng}}, \bibinfo
  {author} {\bibfnamefont {S.}~\bibnamefont {Wang}}, \bibinfo {author}
  {\bibfnamefont {G.}~\bibnamefont {Chen}}, \bibinfo {author} {\bibfnamefont
  {L.}~\bibnamefont {Ma}}, \emph {et~al.},\ }\bibfield  {title} {\bibinfo
  {title} {Gaps induced by inversion symmetry breaking and second-generation
  dirac cones in graphene/hexagonal boron nitride},\ }\href
  {https://www.nature.com/articles/nphys3856} {\bibfield  {journal} {\bibinfo
  {journal} {Nature Physics}\ }\textbf {\bibinfo {volume} {12}},\ \bibinfo
  {pages} {1111} (\bibinfo {year} {2016})}\BibitemShut {NoStop}%
\bibitem [{\citenamefont {Kerelsky}\ \emph {et~al.}(2019)\citenamefont
  {Kerelsky}, \citenamefont {McGilly}, \citenamefont {Kennes}, \citenamefont
  {Xian}, \citenamefont {Yankowitz}, \citenamefont {Chen}, \citenamefont
  {Watanabe}, \citenamefont {Taniguchi}, \citenamefont {Hone}, \citenamefont
  {Dean} \emph {et~al.}}]{kerelsky2019maximized}%
  \BibitemOpen
  \bibfield  {author} {\bibinfo {author} {\bibfnamefont {A.}~\bibnamefont
  {Kerelsky}}, \bibinfo {author} {\bibfnamefont {L.~J.}\ \bibnamefont
  {McGilly}}, \bibinfo {author} {\bibfnamefont {D.~M.}\ \bibnamefont {Kennes}},
  \bibinfo {author} {\bibfnamefont {L.}~\bibnamefont {Xian}}, \bibinfo {author}
  {\bibfnamefont {M.}~\bibnamefont {Yankowitz}}, \bibinfo {author}
  {\bibfnamefont {S.}~\bibnamefont {Chen}}, \bibinfo {author} {\bibfnamefont
  {K.}~\bibnamefont {Watanabe}}, \bibinfo {author} {\bibfnamefont
  {T.}~\bibnamefont {Taniguchi}}, \bibinfo {author} {\bibfnamefont
  {J.}~\bibnamefont {Hone}}, \bibinfo {author} {\bibfnamefont {C.}~\bibnamefont
  {Dean}}, \emph {et~al.},\ }\bibfield  {title} {\bibinfo {title} {Maximized
  electron interactions at the magic angle in twisted bilayer graphene},\
  }\href {https://www.nature.com/articles/s41586-019-1431-9} {\bibfield
  {journal} {\bibinfo  {journal} {Nature}\ }\textbf {\bibinfo {volume} {572}},\
  \bibinfo {pages} {95} (\bibinfo {year} {2019})}\BibitemShut {NoStop}%
\bibitem [{\citenamefont {Ghosal}\ \emph {et~al.}(2022)\citenamefont {Ghosal},
  \citenamefont {Gruschwitz}, \citenamefont {Koch}, \citenamefont {Gemming},\
  and\ \citenamefont {Tegenkamp}}]{PhysRevLett.129.116802}%
  \BibitemOpen
  \bibfield  {author} {\bibinfo {author} {\bibfnamefont {C.}~\bibnamefont
  {Ghosal}}, \bibinfo {author} {\bibfnamefont {M.}~\bibnamefont {Gruschwitz}},
  \bibinfo {author} {\bibfnamefont {J.}~\bibnamefont {Koch}}, \bibinfo {author}
  {\bibfnamefont {S.}~\bibnamefont {Gemming}},\ and\ \bibinfo {author}
  {\bibfnamefont {C.}~\bibnamefont {Tegenkamp}},\ }\bibfield  {title} {\bibinfo
  {title} {Proximity-induced gap opening by twisted plumbene in epitaxial
  graphene},\ }\href {https://doi.org/10.1103/PhysRevLett.129.116802}
  {\bibfield  {journal} {\bibinfo  {journal} {Phys. Rev. Lett.}\ }\textbf
  {\bibinfo {volume} {129}},\ \bibinfo {pages} {116802} (\bibinfo {year}
  {2022})}\BibitemShut {NoStop}%
\bibitem [{\citenamefont {Shi}\ \emph {et~al.}(2021)\citenamefont {Shi},
  \citenamefont {Zhu},\ and\ \citenamefont {MacDonald}}]{PhysRevB.103.075122}%
  \BibitemOpen
  \bibfield  {author} {\bibinfo {author} {\bibfnamefont {J.}~\bibnamefont
  {Shi}}, \bibinfo {author} {\bibfnamefont {J.}~\bibnamefont {Zhu}},\ and\
  \bibinfo {author} {\bibfnamefont {A.~H.}\ \bibnamefont {MacDonald}},\
  }\bibfield  {title} {\bibinfo {title} {Moir\'e commensurability and the
  quantum anomalous hall effect in twisted bilayer graphene on hexagonal boron
  nitride},\ }\href {https://doi.org/10.1103/PhysRevB.103.075122} {\bibfield
  {journal} {\bibinfo  {journal} {Phys. Rev. B}\ }\textbf {\bibinfo {volume}
  {103}},\ \bibinfo {pages} {075122} (\bibinfo {year} {2021})}\BibitemShut
  {NoStop}%
\bibitem [{\citenamefont {Stacey}(1982)}]{PhysRevD.26.468}%
  \BibitemOpen
  \bibfield  {author} {\bibinfo {author} {\bibfnamefont {R.}~\bibnamefont
  {Stacey}},\ }\bibfield  {title} {\bibinfo {title} {Eliminating lattice
  fermion doubling},\ }\href {https://doi.org/10.1103/PhysRevD.26.468}
  {\bibfield  {journal} {\bibinfo  {journal} {Phys. Rev. D}\ }\textbf {\bibinfo
  {volume} {26}},\ \bibinfo {pages} {468} (\bibinfo {year} {1982})}\BibitemShut
  {NoStop}%
\bibitem [{\citenamefont {Bender}\ \emph {et~al.}(1983)\citenamefont {Bender},
  \citenamefont {Milton},\ and\ \citenamefont {Sharp}}]{PhysRevLett.51.1815}%
  \BibitemOpen
  \bibfield  {author} {\bibinfo {author} {\bibfnamefont {C.~M.}\ \bibnamefont
  {Bender}}, \bibinfo {author} {\bibfnamefont {K.~A.}\ \bibnamefont {Milton}},\
  and\ \bibinfo {author} {\bibfnamefont {D.~H.}\ \bibnamefont {Sharp}},\
  }\bibfield  {title} {\bibinfo {title} {Consistent formulation of fermions on
  a minkowski lattice},\ }\href {https://doi.org/10.1103/PhysRevLett.51.1815}
  {\bibfield  {journal} {\bibinfo  {journal} {Phys. Rev. Lett.}\ }\textbf
  {\bibinfo {volume} {51}},\ \bibinfo {pages} {1815} (\bibinfo {year}
  {1983})}\BibitemShut {NoStop}%
\bibitem [{\citenamefont {Borunda}\ \emph {et~al.}(2011)\citenamefont
  {Borunda}, \citenamefont {Berezovsky}, \citenamefont {Westervelt},\ and\
  \citenamefont {Heller}}]{borunda2011imaging}%
  \BibitemOpen
  \bibfield  {author} {\bibinfo {author} {\bibfnamefont {M.~F.}\ \bibnamefont
  {Borunda}}, \bibinfo {author} {\bibfnamefont {J.}~\bibnamefont {Berezovsky}},
  \bibinfo {author} {\bibfnamefont {R.~M.}\ \bibnamefont {Westervelt}},\ and\
  \bibinfo {author} {\bibfnamefont {E.~J.}\ \bibnamefont {Heller}},\ }\bibfield
   {title} {\bibinfo {title} {Imaging universal conductance fluctuations in
  graphene},\ }\href {https://pubs.acs.org/doi/full/10.1021/nn103450d}
  {\bibfield  {journal} {\bibinfo  {journal} {ACS nano}\ }\textbf {\bibinfo
  {volume} {5}},\ \bibinfo {pages} {3622} (\bibinfo {year} {2011})}\BibitemShut
  {NoStop}%
\bibitem [{\citenamefont {Borunda}\ \emph {et~al.}(2013)\citenamefont
  {Borunda}, \citenamefont {Hennig},\ and\ \citenamefont
  {Heller}}]{PhysRevB.88.125415}%
  \BibitemOpen
  \bibfield  {author} {\bibinfo {author} {\bibfnamefont {M.~F.}\ \bibnamefont
  {Borunda}}, \bibinfo {author} {\bibfnamefont {H.}~\bibnamefont {Hennig}},\
  and\ \bibinfo {author} {\bibfnamefont {E.~J.}\ \bibnamefont {Heller}},\
  }\bibfield  {title} {\bibinfo {title} {Ballistic versus diffusive transport
  in graphene},\ }\href {https://doi.org/10.1103/PhysRevB.88.125415} {\bibfield
   {journal} {\bibinfo  {journal} {Phys. Rev. B}\ }\textbf {\bibinfo {volume}
  {88}},\ \bibinfo {pages} {125415} (\bibinfo {year} {2013})}\BibitemShut
  {NoStop}%
\bibitem [{\citenamefont {Beenakker}\ \emph {et~al.}(2023)\citenamefont
  {Beenakker}, \citenamefont {Don{\'\i}s~Vela}, \citenamefont {Lemut},
  \citenamefont {Pacholski},\ and\ \citenamefont
  {Tworzyd{\l}o}}]{beenakker2023tangent}%
  \BibitemOpen
  \bibfield  {author} {\bibinfo {author} {\bibfnamefont {C.}~\bibnamefont
  {Beenakker}}, \bibinfo {author} {\bibfnamefont {A.}~\bibnamefont
  {Don{\'\i}s~Vela}}, \bibinfo {author} {\bibfnamefont {G.}~\bibnamefont
  {Lemut}}, \bibinfo {author} {\bibfnamefont {M.}~\bibnamefont {Pacholski}},\
  and\ \bibinfo {author} {\bibfnamefont {J.}~\bibnamefont {Tworzyd{\l}o}},\
  }\bibfield  {title} {\bibinfo {title} {Tangent fermions: Dirac or majorana
  fermions on a lattice without fermion doubling},\ }\href
  {https://doi.org/10.1002/andp.202300081} {\bibfield  {journal} {\bibinfo
  {journal} {Annalen der Physik}\ ,\ \bibinfo {pages} {2300081}} (\bibinfo
  {year} {2023})}\BibitemShut {NoStop}%
\bibitem [{\citenamefont {Pacholski}\ \emph {et~al.}(2021)\citenamefont
  {Pacholski}, \citenamefont {Lemut}, \citenamefont {Tworzyd{\l}o},\ and\
  \citenamefont {Beenakker}}]{pacholski2021generalized}%
  \BibitemOpen
  \bibfield  {author} {\bibinfo {author} {\bibfnamefont {M.}~\bibnamefont
  {Pacholski}}, \bibinfo {author} {\bibfnamefont {G.}~\bibnamefont {Lemut}},
  \bibinfo {author} {\bibfnamefont {J.}~\bibnamefont {Tworzyd{\l}o}},\ and\
  \bibinfo {author} {\bibfnamefont {C.}~\bibnamefont {Beenakker}},\ }\bibfield
  {title} {\bibinfo {title} {Generalized eigenproblem without fermion doubling
  for dirac fermions on a lattice},\ }\href
  {https://doi.org/10.21468/SciPostPhys.11.6.105} {\bibfield  {journal}
  {\bibinfo  {journal} {SciPost Physics}\ }\textbf {\bibinfo {volume} {11}},\
  \bibinfo {pages} {105} (\bibinfo {year} {2021})}\BibitemShut {NoStop}%
\bibitem [{\citenamefont {Abergel}\ \emph {et~al.}(2010)\citenamefont
  {Abergel}, \citenamefont {Apalkov}, \citenamefont {Berashevich},
  \citenamefont {Ziegler},\ and\ \citenamefont
  {Chakraborty}}]{abergel2010properties}%
  \BibitemOpen
  \bibfield  {author} {\bibinfo {author} {\bibfnamefont {D.}~\bibnamefont
  {Abergel}}, \bibinfo {author} {\bibfnamefont {V.}~\bibnamefont {Apalkov}},
  \bibinfo {author} {\bibfnamefont {J.}~\bibnamefont {Berashevich}}, \bibinfo
  {author} {\bibfnamefont {K.}~\bibnamefont {Ziegler}},\ and\ \bibinfo {author}
  {\bibfnamefont {T.}~\bibnamefont {Chakraborty}},\ }\bibfield  {title}
  {\bibinfo {title} {Properties of graphene: a theoretical perspective},\
  }\href {https://doi.org/10.1080/00018732.2010.487978} {\bibfield  {journal}
  {\bibinfo  {journal} {Advances in Physics}\ }\textbf {\bibinfo {volume}
  {59}},\ \bibinfo {pages} {261} (\bibinfo {year} {2010})}\BibitemShut
  {NoStop}%
\bibitem [{\citenamefont {Ziegler}(2009)}]{PhysRevLett.102.126802}%
  \BibitemOpen
  \bibfield  {author} {\bibinfo {author} {\bibfnamefont {K.}~\bibnamefont
  {Ziegler}},\ }\bibfield  {title} {\bibinfo {title} {Random-gap model for
  graphene and graphene bilayers},\ }\href
  {https://doi.org/10.1103/PhysRevLett.102.126802} {\bibfield  {journal}
  {\bibinfo  {journal} {Phys. Rev. Lett.}\ }\textbf {\bibinfo {volume} {102}},\
  \bibinfo {pages} {126802} (\bibinfo {year} {2009})}\BibitemShut {NoStop}%
\bibitem [{\citenamefont {Ziegler}\ and\ \citenamefont
  {Sinner}(2010)}]{PhysRevB.81.241404}%
  \BibitemOpen
  \bibfield  {author} {\bibinfo {author} {\bibfnamefont {K.}~\bibnamefont
  {Ziegler}}\ and\ \bibinfo {author} {\bibfnamefont {A.}~\bibnamefont
  {Sinner}},\ }\bibfield  {title} {\bibinfo {title} {Transport in finite
  graphene samples with a random gap},\ }\href
  {https://doi.org/10.1103/PhysRevB.81.241404} {\bibfield  {journal} {\bibinfo
  {journal} {Phys. Rev. B}\ }\textbf {\bibinfo {volume} {81}},\ \bibinfo
  {pages} {241404} (\bibinfo {year} {2010})}\BibitemShut {NoStop}%
\end{thebibliography}%

\end{document}